\pdfoutput=1

\documentclass[twocolumn,superscriptaddress,aps,preprintnumbers,amsmath,amssymb,prd,nofootinbib]{revtex4}

\usepackage{epsfig}
\usepackage{amsmath}
\usepackage{amssymb}
\usepackage{subfigure}
\usepackage{cancel}
\usepackage{textcomp}
\usepackage{calc}
\usepackage{graphicx}
\usepackage{dcolumn}
\usepackage[dvipsnames,table,xcdraw]{xcolor}
\usepackage{pifont}
\usepackage{slashed}

\usepackage{xcolor}
\RequirePackage[normalem]{ulem}

\RequirePackage[ colorlinks=true,urlcolor=blue,anchorcolor=blue,citecolor=blue,filecolor=blue,
               linkcolor=blue,menucolor=blue,linktocpage=true,pdfproducer=medialab,pagebackref=true]{hyperref}
 \hypersetup{
 colorlinks   = true, 
 urlcolor     = PineGreen, 
 linkcolor    = blue, 
  citecolor   = blue 
}
\usepackage[capitalise]{cleveref}
\usepackage[normalem]{ulem}
\usepackage{orcidlink}

\newcommand\ga{g_{a \gamma \gamma}}

\newcommand{\axions}{\text{ALPs}}
\newcommand{\axion}{\text{ALP}}

\newcommand{\Axion}{\text{ALP}}

\begin{document}


\preprint{DESY-22-064}
\title{$\Axion$ Searches at the LHC: FASER as a Light Shining through Walls Experiment}

\author{Felix Kling \orcidlink{0000-0002-3100-6144}}
\email{felix.kling@desy.de}
\affiliation{Deutsches Elektronen-Synchrotron DESY, Germany}
\author{Pablo Qu\'ilez \orcidlink{0000-0002-4327-2706}}
\email{pablo.quilez@desy.de}
\affiliation{Deutsches Elektronen-Synchrotron DESY, Germany}


\begin{abstract}
We propose the use of FASER as a light-shining-through-walls experiment to search for axions and axion-like particles (ALPs). LHC collisions generate a high intensity and high energy photon flux in the forward direction which can oscillate into ALPs in the magnetic fields that are used to confine the beam. These $\axions$ then pass through about 100~m of rock before reaching the magnetic fields of FASER, where they can convert back into photons and be detected by an electromagnetic calorimeter. In the next years, FASER and its successor FASER2 at the Forward Physics Facility will be able to explore regions of the $\axion$ parameter space inaccessible by other laboratory-based experiments. 
\end{abstract}


\maketitle


\section{Introduction}

Axions are among the best motivated scenarios of new physics beyond the Standard Model (BSM). Most famously, the QCD axion arises as a low-energy consequence of the elegant Peccei-Quinn solution to the strong CP problem of the SM~\cite{Peccei:1977hh, Kim:1979if, Shifman:1979if, Zhitnitsky:1980tq, Dine:1981rt,Kim:1984pt}. But the interest in axion-like particles 
goes well beyond the QCD axion. Indeed these pseudo-Nambu–Goldstone bosons arise in a plethora of BSM constructions such as theories with extra dimensions~\cite{Dienes:1999gw}, Majoron models~\cite{Gelmini:1980re}, theories of flavor~\cite{Davidson:1981zd,Wilczek:1982rv} and are ubiquitous within string theory~\cite{Cicoli:2013ana}. They all share the basic properties that stem from its pseudo Nambu-Goldstone nature: axions are light and present feeble derivative or anomalous interactions suppressed by the axion decay constant $f_a$. Moreover, they are excellent dark matter candidates~\cite{Abbott:1982af, Dine:1982ah, Preskill:1982cy}. 

As a consequence, a strong experimental program has been developed and in the next decades we will explore large regions of the $\axion$ parameter space. Most of the experiments only take advantage of the $\axion$ coupling to photons. In this case the relevant Lagrangian reads 
\begin{align}
    \mathcal{L} \supset - \frac{1}{2} m_a^2 a^2 -\frac{1}{4} \ga a F_{\mu \nu} \tilde{F}^{\mu \nu}
\end{align}
where $F_{\mu \nu}$ is he electromagnetic field strength and  $\tilde{F}^{\mu \nu}$ its dual. The resulting parameter space is therefore spanned by the $\axion$ mass $m_a$ and the $\axion$ coupling $\ga$. $\Axion$ search experiments utilize a variety of experimental techniques that range from laboratory-based light-shining-through walls (LSW) experiments such as ALPS II~\cite{Spector:2019ooq}, helioscopes looking for solar $\axions$ such as CAST~\cite{Irastorza:2013dav, Anastassopoulos:2017ftl} and the future IAXO~\cite{Armengaud:2014gea, Giannotti:2016drd}, haloscopes looking for dark matter $\axions$~\cite{Majorovits:2017ppy, Budker:2013hfa, Ruoso:2015ytk, Experiment:2017icw, Zenner:2013rta, Adams:2022pbo} but also colliders~\cite{Bauer:2017ris, Alonso-Alvarez:2018irt}, flavor probes~\cite{Izaguirre:2016dfi, CidVidal:2018blh, Gavela:2019wzg, Bauer:2021mvw}, as well as beam dump~\cite{Dobrich:2019dxc} and neutrino~\cite{Brdar:2020dpr, Coloma:2022hlv} experiments for larger masses. 
There are also a number of astrophysical constraints (typically through energy-loss arguments) and cosmological bounds which provide a nice complementarity with respect to proposed experiments. However, these considerations vary in their robustness. 
For example, the supernova bound from the SN1987 neutrino signal~\cite{Raffelt:1990yz} has been recently put in question~\cite{Bar:2019ifz}
although not everyone in the community agrees with this criticism and it has been argued that the SN1987 also constrains the $\axion$ through the $\gamma$-ray signal and the diffuse cosmic $\gamma$-ray background~\cite{Caputo:2021rux}. Similarly some cosmological bounds can be significantly relaxed in non-standard cosmological scenarios~\cite{Depta:2020wmr, Carenza:2020zil}. This is why there is a growing community interest in purely laboratory based $\axion$ experiments, which has been advocated for (e.g. in Ref.~\cite{Jaeckel:2006xm}). 

In this letter we contribute to this effort by noting that the LHC accelerator in combination with the recently installed FASER experiment~\cite{Feng:2017uoz, FASER:2018ceo, FASER:2018bac} can be used as a LSW experiment without requiring any modification of the current experimental setup. The LSW detection principle is based on the coherent $\axion$-photon oscillation in the presence of an external magnetic field. A strong photon beam is directed at a thick wall with a photon detector placed behind it. In the absence of $\axions$, no signal is expected in the detector. However, if $\axions$ exist the strong magnetic fields perpendicular to the beam located at both sides of the wall may induce the photons to convert to $\axions$ which can freely pass through the wall. These $\axions$ can then oscillate back to photons and be detected on the other side of the wall. Effectively, this double conversion mechanism allows a small fraction of the photon beam to pass through the wall where it can be detected. LSW experiments use this mechanism to search for $\axions$.

\begin{figure*}[t]
    \centering
    \includegraphics[width=0.98\textwidth]{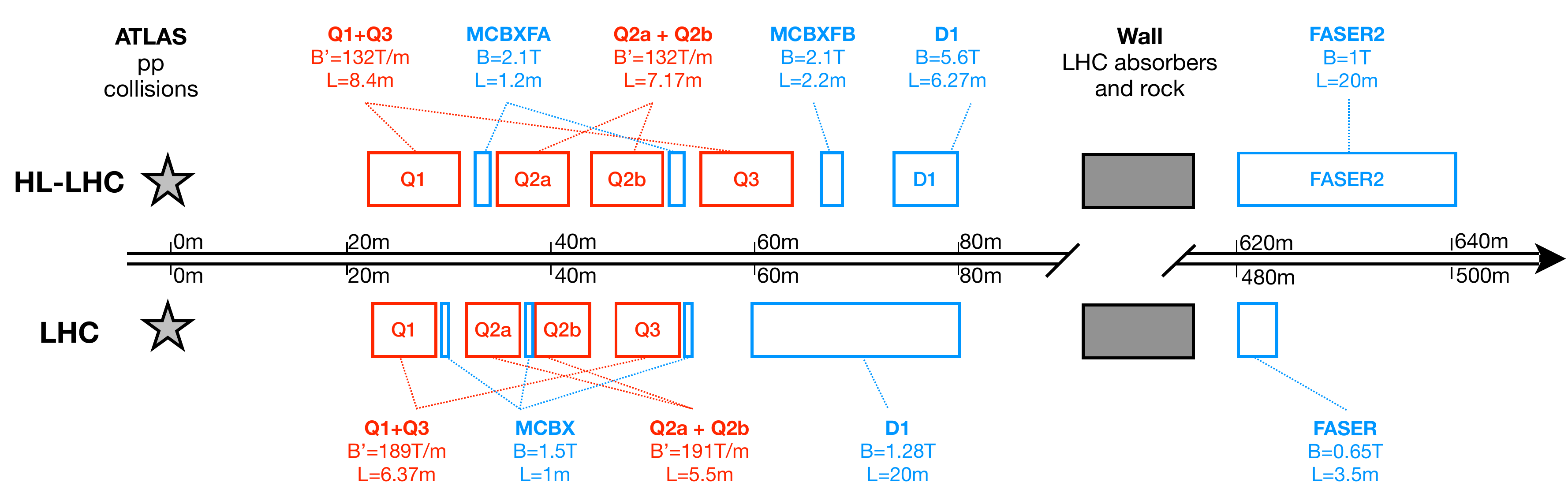}
    \caption{Considered magnetic fields at the HL-LHC (top) and LHC Run 3 (bottom) downstream from the ATLAS collisions point. We show dipoles in blue and quadruples in red, and indicate their corresponding length $L$, magnetic field $B$ and field gradient $B'$. Here we have used the magnetic fields as presented in Refs.~\cite{Kling:2021gos} and \cite{ZurbanoFernandez:2020cco}.}
    \label{fig:layout}
\end{figure*}

In this study we explore the sensitivity of both FASER and future forward detectors proposed in the context of the Forward Physics Facility (FPF)~\cite{Anchordoqui:2021ghd, Feng:2022inv} for light $\axion$ searches. The main advantage of this proposal with respect to existing LSW experiments lies on the high energies of LHC photons which allows us to explore higher $\axion$ masses. With respect to other accelerator searches, this proposal enjoys a better sensitivity at lower masses since it relies on coherent conversion in an external macroscopic magnetic field instead of production/detection through scattering or perturbative decay. In the following, we will discuss the production of $\axions$ at the LHC in \cref{AxionProduction}, their detection in \cref{AxionDetection}, and the resulting sensitivity in \cref{Results}.

\section{$\Axion$ Production at the LHC}
\label{AxionProduction}

Since more than half a century, hadron colliders have been the primary tools to probe the fundamental laws of nature at the highest energies. The currently most energetic realization of these hadron collider experiments is the LHC at CERN, which started to collect data in 2010 and is expected to run until 2040. With a collision energy of 14~TeV, it is able to produce heavy particles with masses at the weak scale and beyond, and large detectors like ATLAS and CMS have been build around the LHC collision point to detect their decay products. Their biggest achievement thus far is the discovery of the Higgs boson in 2012~\cite{ATLAS:2012yve, CMS:2012qbp} and the main objectives for their remaining operation are to study the properties of the Higgs and to search for new heavy particles at the TeV scale. In addition, the LHC also produces an enormous number of energetic light particles, which are mainly produced in the forward direction. As first pointed out by the authors of Ref.~\cite{Feng:2017uoz}, this provides an additional opportunity to search for light and weakly coupled particles. In the following, we will apply  this idea in the context of $\axions$.

\begin{figure}[t]
    \centering
    \includegraphics[width=0.495\textwidth]{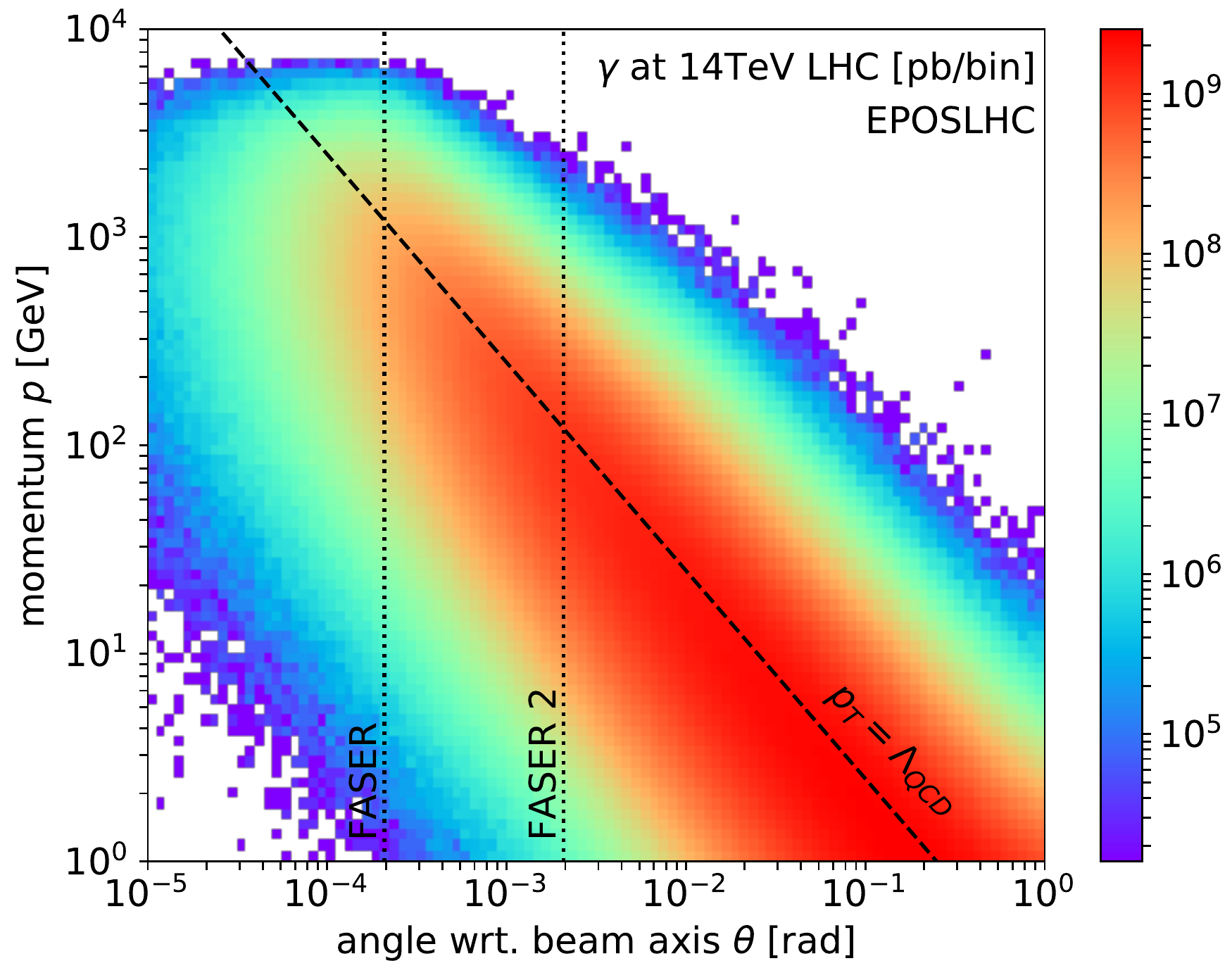}
    \caption{
    Differential production rate of photons at 14 TeV LHC predicted by \texttt{EPOS-LHC} in the ($\theta$,p) plane, where $\theta$ and p are the proton's angle with respect to the beam axis and momentum, respectively. The angular coverage for FASER and FASER 2 are indicated by the vertical dotted lines. 
    }
    \label{fig:distribution}
\end{figure}

As a byproduct of proton-proton collisions, the ATLAS experiment also produces an intense and strongly collimated beam of photons with energy $E_\gamma\sim\text{TeV}$ in the far forward direction. These photons mainly result from the decay of neutral pions, which were for example formed in the hadronization of the beam remnants. In this study, we use the double differential photon spectrum $d^2\sigma/dE d\theta$ (with respect to the photon energy and angle with respect to the beam axis) provided by the 
\href{https://github.com/KlingFelix/FORESEE}{\color{black} \texttt{FORESEE}}~\cite{Kling:2021fwx} tool, which were obtained by the \texttt{EPOS-LHC}~\cite{Pierog:2013ria} event generator as implemented in the \texttt{CRMC}~\cite{CRMC} package. The photon distribution is shown in \cref{fig:distribution}. Roughly speaking, we expect about $10^{14}$ photons per fb$^{-1}$ of integrated luminosity to be produced within 1~mrad of the beam axis.  

These photons travel down the beam pipe, where they pass through a variety of magnets before being absorbed in the LHC infrastructure. An overview of the magnetic fields that are relevant for our study is presented in Fig.~\ref{fig:layout}. This includes the so called inner triplet, which is a set of quadrupole magnets to focus the proton beam before the collision, and the D1 dipole magnet, which separates the incoming and outgoing proton beams before they enter their separate beam pipes. Note that the exact dimensions and strength of these magnets will undergo some changes in the future. The LHC in its current configuration will operate until about 2025. Afterwards, it will be upgraded in preparation of the following high-luminosity LHC (HL-LHC) era. When modelling the magnetic fields, we assume that the dipole magnets have a constant field strength $B$,  while quadrupole magnets are specified by a field gradient $B'$ and their field strength is given by $B= B' \cdot r$ where $r$ denotes the distance from the center~\cite{Nevay:2018zhp}.
We further model the beam pipe as a cylinder with diameter of 5.3~cm and assume that photons are quickly absorbed once they leave the vacuum.

When passing these strong magnetic fields, the photons could oscillate into $\axions$. The corresponding conversion probability can be written as~\cite{Redondo:2010dp}
\begin{equation}
    P_{\gamma \to a} \simeq \frac{1}{2} \times \frac{\ga^2}{4} \left| \int dz \, \vec B(z) \, e^{i q z} \right|^2  ,
\end{equation}
where we integrate the magnetic field coherently along the trajectory of the photon. The additional factor $1/2$ in front accounts for the fact that the photon beam is unpolarized and only the photons with a polarization parallel to the magnetic field can convert into $\axions$ since $F\widetilde F\sim\vec E\cdot\vec B$. Here $q=E_\gamma-(E_\gamma^{2}-m_{a}^{2})^{1/2}\simeq m_a^2/(2E_\gamma)$ is the momentum transfer. In our case, the magnetic fields along a single photon's trajectory are piece-wise constant which allows us to write 
\begin{equation}
    P_{\gamma \to a} \simeq  \frac{\ga^2}{8} \left| \sum_i \vec B_i\, L_i\,  e^{iq z_i}\frac{\sin(q L_i /2)}{qL_i/2} \right|^2  . 
    \label{Eq:conversionProb}
\end{equation}
where $\vec B_i$ is the field strength of the dipole and quadrupole magnets seen by the photon, $L_i$ is the length and $z_i$ is the central position of each magnet. Using that the typical length of the magnets is $L \sim 100$~m and the typical energy of the photons is $E_\gamma \sim 1$~TeV, we can distinguish two relevant kinematic regimes. For small masses $m_a \ll \mathcal{O}(100)$~eV and hence $qL \ll 1$, the conversion takes place coherently along the full length of the magnetic field. This can be understood from the fact that the phase velocity of both the $\axion$ and the photon are approximately the same and thus the probability amplitude of the conversion at any point interferes constructively with that in the rest of the conversion region. As a consequence of the probability is enhanced, scaling with the square of the length $\propto L^2$ and becomes mass independent. However, for large masses $m_a \gg  \mathcal{O}(100)$~eV and hence $qL \gg 1$ the $\axion$ and photon states do not remain in phase along all the conversion length. Hence, the probability becomes suppressed $P_{\gamma \to a} \sim 1/m_a^4$ and other production mechanisms become more dominant.

Eventually, the photons produced at the LHC will collide with elements of the LHC infrastructure and be absorbed. For very forward going photons, this occurs in the TAN neutral particle absorber, which is located at about 140~m downstream from the ATLAS interaction point. This provides a further opportunity for $\axion$ production via the Primakoff process, in which the photon converts
into an axion when interacting with nuclei in the target material. In the limit of small $\axion$ masses $m_a$ and small momentum transfer $q$, the corresponding cross section can be written as~\cite{Primakoff:1951iae, Tsai:1986tx, Feng:2018pew}
\begin{equation}
    \sigma_\text{Prim} \simeq  \frac{\ga^2}{2} \alpha Z^2 \! \int \! F^2(q^2)\,  d\log q = 9.8 \times \ga^2 .
    \label{Eq:Primxsec}
\end{equation}
Here we have assumed that the target material is made of iron with atomic number $Z=26$ and integrated the form factor $F(q^2)$ as in used in Refs.~\cite{Tsai:1986tx, Feng:2018pew}. The photon to $\axion$ conversion probability is given by 
\begin{equation}
    P_{\gamma \to a} =  \frac{ \sigma_\text{Prim} }{ \sigma_\text{conv} } = 7.4\cdot10^{-4} \times (\ga \cdot \text{GeV})^2 , 
\end{equation}
where $\sigma_\text{conv} \simeq5$~barn is the conventional cross section for the photon interaction with the target material (dominated by pair-production). At the LHC, this production mechanism dominates for high $\axion$ masses $m_a > 1$~keV. 

After the TAN neutral particle absorber, essentially all photons originating from the primary collision have been absorbed. The produced $\axions$ will then continue to propagate along the beam collision axis. Roughly 270~m downstream from the ATLAS interaction point, the LHC magnets start to deflect the beam to bring it on its roughly circular orbit and LHC tunnels curve away from the beam collision axis.

\section{Axion Detection}
\label{AxionDetection}
To detect the produced beam of ultrarelativistic $\axions$ an additional magnetic field is needed to convert them back into photons. As we will see, the recently installed FASER experiment is already placed directly in the beam of LHC $\axions$ and equipped with a strong magnetic field to perform that task.

About 480~m downstream from ATLAS, the beam collision axis intersects with the service tunnel TI12. This location provides a unique opportunity to access the beam of light and weakly interacting particles that are produced in the forward direction of the LHC~\cite{Feng:2017uoz}. The FASER experiment has recently been installed there to exploit this opportunity~\cite{Feng:2017uoz, FASER:2018ceo, FASER:2018bac, FASER:2019aik, FASER:2021cpr, FASER:2021ljd}. The main goal of FASER search for light long-lived particles at the LHC~\cite{Feng:2017uoz, Feng:2017vli, Kling:2018wct, Feng:2018noy, FASER:2018eoc}. In addition, the experiment also contains the FASER$\nu$ neutrino detector which will for the first time detect and study of TeV energy neutrinos produced at the LHC~\cite{FASER:2019dxq, FASER:2020gpr, FASER:2021mtu}. The FASER experiment will operate during LHC Run~3, from 2022 to 2025, with an expected integrated luminosity of 150~fb$^{-1}$.

The experiment consists of a decay volume and tracking spectrometer which are placed inside an otherwise empty magnetized area of length of 3.5~m, diameter of 20~cm and field strength of 0.6~T. Located behind the spectrometer is a preshower detector~\cite{Boyd:2803084} for photon identification and an electromagnetic calorimeter to measure the photon energy. FASER therefore already provides an ideal setup for $\axion$ detection. The corresponding probability of $\axion$ to photon conversion can be written as
\begin{equation}
    \!\!\!P_{a \to \gamma} \!\simeq\! \frac{\ga^2}{4} \left| \int\!\! dz B(z) e^{i q z} \right|^2 \!\!\!=\! \frac{\ga^2}{4} B^2 L^2 \frac{\sin^2(q L /2)}{(qL/2)^2}.
    \label{Eq:axionphoton conversion probability}
\end{equation}
Similarly to the production, the conversion probability is mass independent for small $\axion$ masses and decreases for $m_a \gg 100$~eV. The expected signal then would be a single high energy photon in the preshower detector and electromagnetic calorimeter\footnote{An interesting characteristic of the photons produced via $a \to \gamma$ conversion in  FASER is that they are linearly polarized in the direction of the magnetic field since $F\widetilde F\sim\vec E\cdot\vec B$. However, this feature is not experimentally accessible at FASER and thus FASER cannot distinguish between a scalar and a pseudoscalar.}.

A continuation of FASER's physics program during the HL-LHC era has been proposed in the context of the FPF~\cite{Anchordoqui:2021ghd, Feng:2022inv}. The FPF would be a new cavern located about 620~m downstream from ATLAS where it will provide space for a variety of experiments and collect 3~ab$^{-1}$ of data. In particular, this would include a larger version of the FASER detector, called FASER2. This detector is envision to be roughly 20~m long with an aperture of 2~m and a magnetic field of 1~T which would increase the $\axion$ to photon conversion probability by roughly two orders of magnitude in comparison to FASER.

\begin{figure*}
    \centering
    \includegraphics[width=0.49\textwidth]{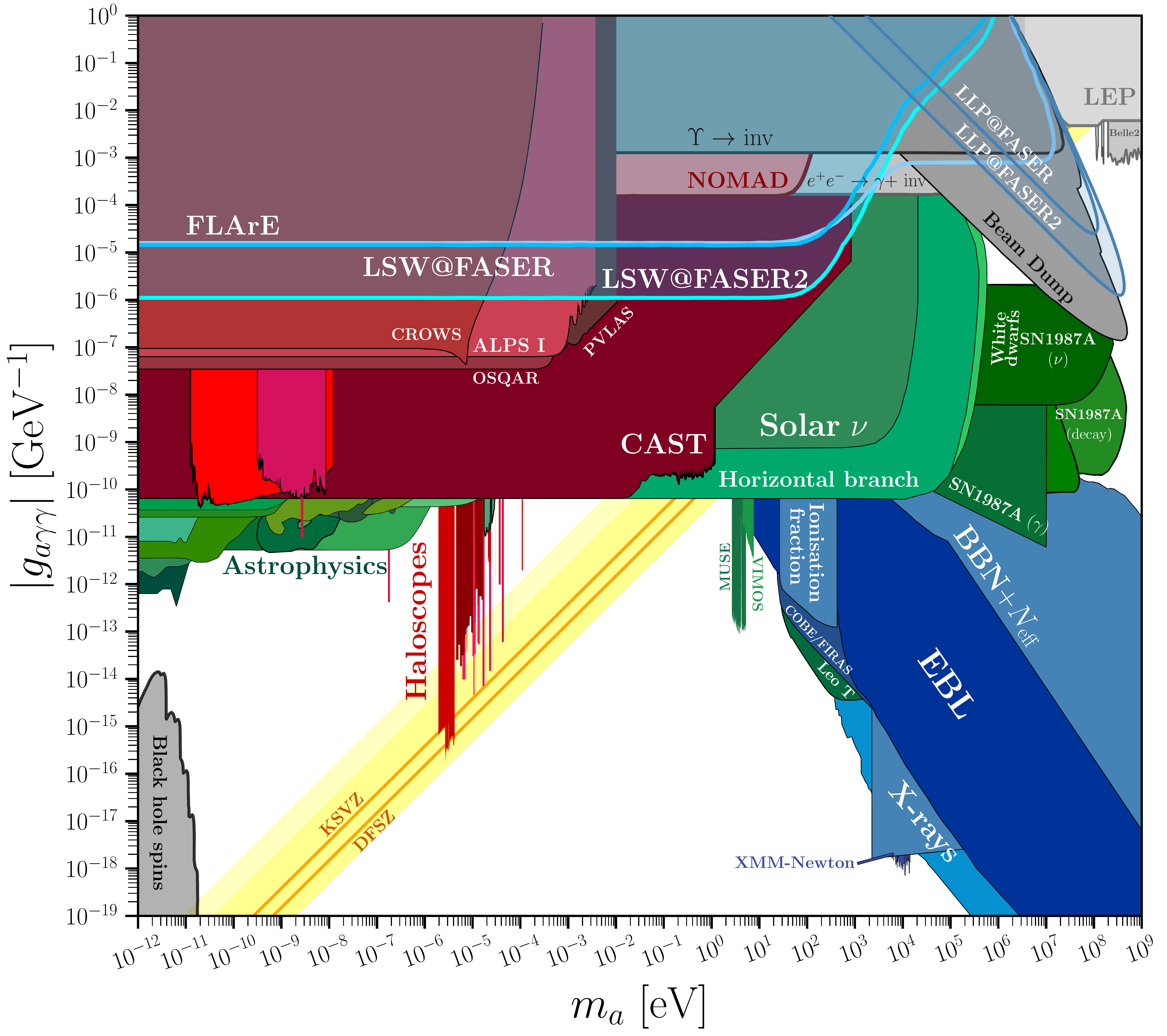}
    \includegraphics[width=0.49\textwidth]{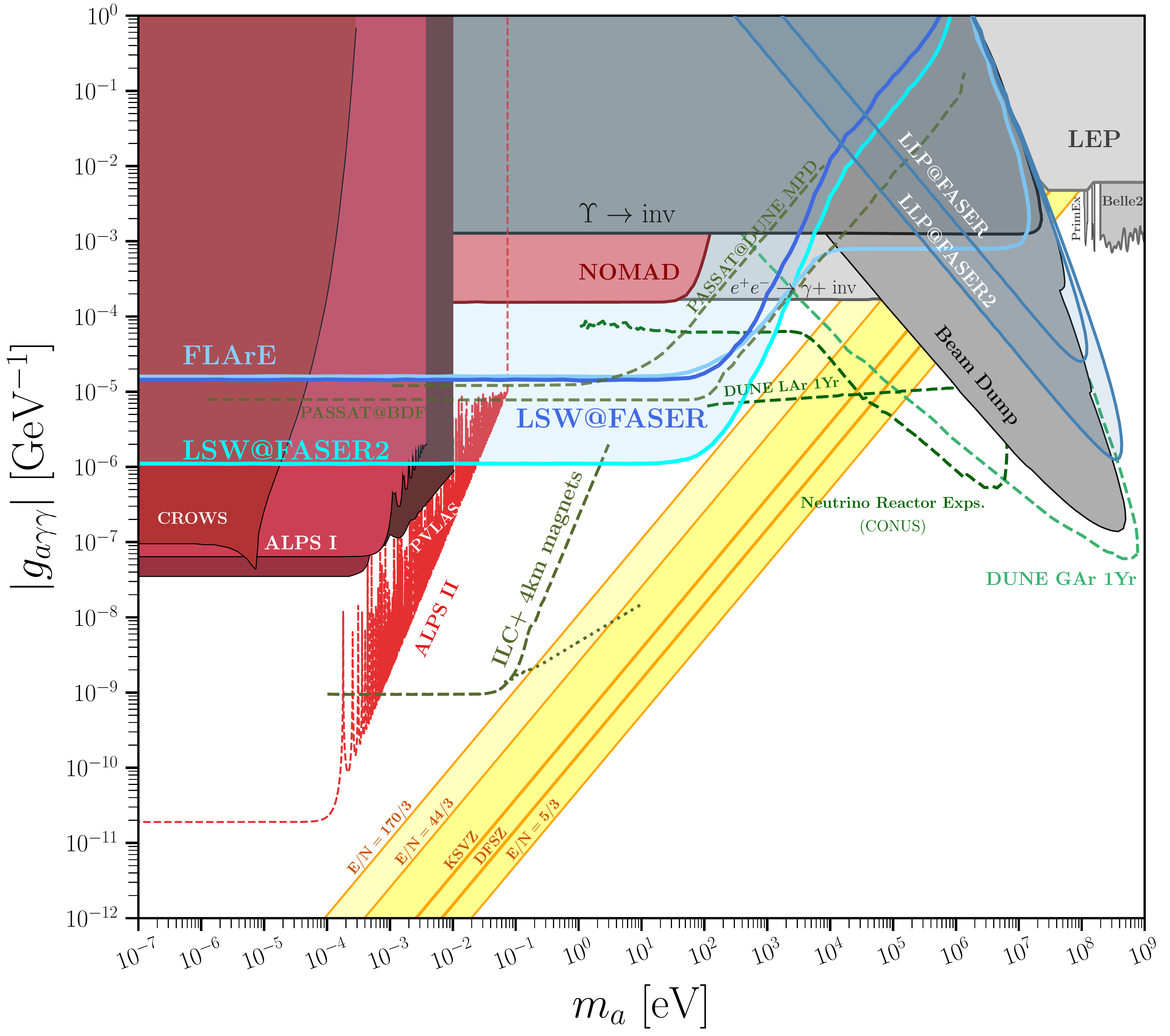}
    \caption{Sensitivity projections of FASER, FASER2 and FLArE as a LSW experiment for $\axion$ searches (shown as blue solid lines) as compared with all current $\axion$ bounds (left) and only purely laboratory based bounds and projections (right). The theoretical predictions for different QCD axion models are shown in orange where $\ga=\alpha / (2 \pi f_{a}) (E/N-1.92)$~\cite{GrillidiCortona:2015jxo}. See text for details. We provide the data of the sensitivity projections of this work as arXiv ancillary files. Existing experimental bounds adapted from \href{https://cajohare.github.io/AxionLimits/}{\color{black} \texttt{AxionLimits}}~\cite{AxionLimits}.}
    \label{fig:PhenoPlots}
\end{figure*}

In addition to FASER2, the FPF contains three neutrino detectors: the emulsion based neutrino detector FASER$\nu$2, the electronic neutrino detector AdvSND and the noble gas neutrino detector FLArE. These could provide an additional opportunity for $\axion$ detection, as these could also undergo the inverse Primakoff scattering process and convert into photons via collisions with the detector material. The signal would be a single highly energetic photon that emerges inside the detector. Perhaps the best suited detector for this purpose is FLArE as it provides timing capabilities, a high spatial resolution, particle identification capabilities as well as a good energy estimate. In particular the particle identification would allow to distinguish photons and electrons, and hence suppress a possible background from neutrino electron scattering that would also provide an isolated electromagnetic shower. The FLArE detector is envisioned to have a 1m$\times$1m$\times$7m target volume filled with liquid argon. The corresponding $\axion$ conversion probability is given by
\begin{equation}
    P_{a \to \gamma} =  \sigma_\text{Prim} n_{Ar} L =  0.028 \times (\ga \cdot \text{GeV})^2 , 
\end{equation}
where $n$ is the number density of liquid argon nuclei and $\sigma_\text{Prim}$ is the cross section for the Primakoff process defined in \cref{Eq:Primxsec}, which has now been evaluated for $Z=18$. 

\section{Results and Discussion}
\label{Results}

We can now combine our previous results to estimate the sensitivity of the LSW setup at the LHC to probe $\axions$\footnote{A notebook containing this analysis has been added to the \href{https://github.com/KlingFelix/FORESEE/blob/main/Models/ALP-LSW/ALP-LSW.ipynb}{\color{black} \texttt{FORESEE}} package}. For this, we compute the expected number of events as
\begin{equation}
    N = \mathcal{L} \int \! dE \int \! d\theta \ \frac{d^2\sigma}{dE d\theta} \cdot P_{\gamma \to a}(E) \cdot P_{a \to \gamma}(E)\,,
\end{equation}
where $\mathcal{L}$ is the LHC's integrated luminosity, $d^2\sigma/(dE d\theta)$ is the double differential photon production cross section with respect to the photon energy and and angle with respect to the beam axis, $P_{\gamma \to a}$ is the probability of photon to $\axion$ conversion in the LHC infrastructure and $P_{a \to \gamma}$ is the probability of $\axion$ to photon conversion in the detector. The angular integral is performed within the detector's angular acceptance. In order to obtain the sensitivity contours we require the number of events to be larger than 3 over the full run time of the experiment for a point in the parameter space to be excluded at 95\% confidence level. Here we use that the converted photons can be identified by FASER's electromagnetic calorimeter and preshower detector and assume that backgrounds can be reduced to a negligible level by the experiment's veto system. A more detailed discussion of possible background sources for the mono-photon signature and how they can be rejected can be found in Ref.~\cite{Jodlowski:2020vhr} in the context of sterile neutrino decays $N \to \nu \gamma$ arising in neutrino dipole portal models.  

We present our obtained sensitivity on the $\axion$ parameter space in \cref{fig:PhenoPlots} alongside existing constraints. In the left panel, we show the broad picture including all existing constraints on $\axions$ arising from helioscopes~\cite{CAST:2007jps, CAST:2017uph, Shilon:2013xma, Ge:2020zww}, haloscopes~\cite{Ouellet:2018beu, Salemi:2021gck, Asztalos2010, ADMX:2018gho, ADMX:2019uok, ADMX:2021nhd, ADMX:2018ogs, Bartram:2021ysp, Crisosto:2019fcj, Lee:2020cfj, Jeong:2020cwz, CAPP:2020utb, Devlin:2021fpq, Grenet:2021vbb, HAYSTAC:2018rwy, HAYSTAC:2020kwv, McAllister:2017lkb, Alesini:2019ajt, Alesini:2020vny, CAST:2020rlf, DePanfilis, Gramolin:2020ict, Arza:2021rrm, Hagmann, Thomson:2019aht} and LSW experiments~\cite{Ehret:2010mh, Betz:2013dza, OSQAR:2015qdv, DellaValle:2015xxa, SAPPHIRES:2021vkz, Ortiz:2020tgs} (red); astrophysics\footnote{The supernova bounds are taken from Refs.~\cite{Lucente:2020whw, Caputo:2021rux, Caputo:2022mah} and should be considered conservative regarding the upper end of the bound that corresponds to the trapping limit~\cite{Caputo:2021rux}.}~\cite{Xiao:2020pra, Keller:2021zbl, Foster:2022fxn, Chan:2021gjl, Wouters:2013hua, Marsh:2017yvc, Reynolds:2019uqt, Reynes:2021bpe, Bolliet:2020ofj, Calore:2021hhn, Calore:2020tjw, Buen-Abad:2020zbd, Fermi-LAT:2016nkz, Meyer:2020vzy, Jacobsen:2022swa, HESS:2013udx, Ayala:2014pea, Wadekar:2021qae, Dessert:2022yqq, Li:2020pcn, Foster:2020pgt, Darling:2020uyo, Battye:2021yue,Vinyoles2015,Jaeckel:2017tud, Payez:2014xsa, Caputo:2021rux, Dessert:2020lil, Blout:2000uc, Regis:2020fhw, Grin:2006aw, Meyer:2016wrm, Thorpe-Morgan:2020rwc, Dekker:2021bos, Dolan:2021rya, Foster:2021ngm} (green); cosmology~\cite{Cadamuro:2011fd, Depta:2020wmr} (blue); and accelerator experiments~\cite{NOMAD:2000usb, Jaeckel:2015jla, DELPHI:1991emv, DELPHI:1994mra, L3:1994shn, L3:1995nbq, CHARM:1985anb, Riordan:1987aw, Dobrich:2017gcm, Bjorken:1988as, Dolan:2017osp, Blumlein:1990ay, Dobrich:2019dxc, NA64:2020qwq, Belle-II:2020jti, Aloni:2019ruo} (gray). The projected sensitivity for searches at FASER and future experiments at the FPF is indicated by the blue solid lines. Even though the regions that can be explored by LSW searches at FASER and FASER2 are excluded by helioscopes, astrophysical or cosmological arguments, these searches will be able to probe regions of the parameter space that have not been explored by any purely laboratory based experiment. This is shown in the right panel, where we show the bounds from current ground-based experiments together with the projections of future experiments and proposals. Note that these experiments are based on different $\axion$ detection strategies which do not rely on extra astrophysical or cosmological assumptions and thus can be considered specially robust. Classical LSW experiments such as ALPS~\cite{Ehret:2010mh}, PVLAS~\cite{DellaValle:2015xxa}, CROWS~\cite{Betz:2013dza} and OSQAR~\cite{OSQAR:2015qdv} as well as a search at NOMAD~\cite{NOMAD:2000usb} utilize the coherent conversion of $\axions$ into photons in an external magnetic field and are shown in redish tones. In contrast, many high energy physics experiment look for the $\axion$ decay into photons and thus are sensitive to heavier $\axions$ with larger couplings, as shown in grey tones. This includes both searches for long-lived $\axions$ at beam dump experiments such as CHARM~\cite{CHARM:1985anb}, E141~\cite{Riordan:1987aw, Dobrich:2017gcm}, E137~\cite{Bjorken:1988as, Dolan:2017osp}, NuCal~\cite{Blumlein:1990ay, Blumlein:1991xh, Dobrich:2019dxc} and  NA64~\cite{NA64:2020qwq} as well as resonance searches for promptly decaying $\axions$ at Belle~2~\cite{Belle-II:2020jti}, PrimEx~\cite{Aloni:2019ruo} and LEP~\cite{Jaeckel:2015jla, DELPHI:1991emv, DELPHI:1994mra, L3:1994shn, L3:1995nbq}. 

As explained in the previous section, our LSW proposal at the LHC benefits from both the coherent conversion in the external magnetic fields and the Primakoff conversion via scattering with nuclei in the target material. The LSW@FASER line corresponds to the reach of the currently installed FASER detector, while LSW@FASER2 represents the reach that will be achieved by the future FASER2 detector at the FPF. 
Similarly to other LWS experiments it is easy to understand the shape of the bound. For small $\axion$ masses, $m_a \ll \mathcal{O}(100)$~eV the factor $qL \ll 1$ and the conversion probability in \cref{Eq:conversionProb} becomes independent on the $\axion$ mass giving rise to a flat bound on $g_{a\gamma\gamma}$. For higher $\axion$ masses, $qL \gg 1$, the coherence is lost and the probability is suppressed by $1/m_a^8$ (the suppression factor $1/q^2\sim E_{\gamma}^2/m_a^4$ arises both at production and detection). The bound on $g_{a\gamma\gamma}$ thus grows as $\propto m_a^2$. For even larger masses, the Primakoff conversion in nuclei becomes the dominant $\axion$ production channel, therefore one of the suppression factors disappears and the bound on $g_{a\gamma\gamma}$ grows only as $\propto m_a$. Furthermore, we also show the projections for the neutrino detector FLArE that will be installed as part of the FPF and could detect the $\axions$ via Primakoff conversion in nuclei and becomes dominant for larger masses. At large masses $m_a \gtrsim 10$~keV, both the production and detection proceeds through the Primakoff process and the sensitivity in $g_{a\gamma\gamma}$ becomes independent of the $\axion$ mass. At high masses and large couplings the $\axion$ lifetime becomes so small that the majority of $\axions$ decay before reaching the detector, hence limiting the reach at FLArE. $\Axion$ decays within the detector volume also provide an additional search channel that has been investigated in Ref.~\cite{Feng:2018noy, FASER:2018eoc}. The corresponding reach of these long-lived particle searches is shown by the LLP@FASER and LLP@FASER2 lines. 

For comparison, we also show the projected sensitivity of other proposed searches and experiments for $\axions$ as dashed lines in \cref{fig:PhenoPlots}. Using a high intensity LASER beam and optical resonant cavities, the ALPS II experiment~\cite{Ortiz:2020tgs} plans to probe $\axion$ masses smaller than $\sim 10^{-4}\,{\rm eV}$ with exceptional sensitivity down to couplings of $\sim 2\times 10^{-11}\,{\rm GeV}^{-1}$. A proposal to probe higher masses has been recently put forward in Ref.~\cite{Fukuda:2022not} for the International Linear Collider (ILC)~\cite{Behnke:2013xla} using the photon beam that would be required for positron production followed by 4km of 1T magnets (adding up conversion and re-conversion lengths). As for the configuration of the magnetic field, on top of the uniform case they also consider non-homogeneous (or \emph{wiggled}) magnetic field profiles which improve the sensitivity for higher $\axion$ masses $m_a\gtrsim0.1{\rm eV}$ (shown with a dotted line).  In addition, the authors of Refs.~\cite{Bonivento:2019sri, Dev:2021ofc} have proposed a class of experiments dubbed PASSAT which use helioscope magnets as detectors for $\axions$ produced via Primakoff effect at the target of beam dump experiments. In \cref{fig:PhenoPlots}, we show two realizations of this idea, one using the CAST magnets placed at the proposed beam dump facility (BDF) and one using the multi-purpose detector (MPD) of the proposed DUNE experiment. To probe $\axions$ with even higher masses, several searches for their interactions via inverse Primakoff scattering as well as their decays at neutrino experiments have been proposed. We show the corresponding sensitivity both for CONUS as an example of coherent elastic neutrino scattering experiments at nuclear reactors~\cite{Dent:2019ueq} as well as DUNE as example of high energy accelerator experiments~\cite{Brdar:2020dpr}. Further searches for the decay of long-lived $\axions$ have also been proposed for other beam dump type experiments, see Ref.~\cite{Dobrich:2019dxc} or Ref.~\cite{Beacham:2019nyx} for recent reviews. 

Regarding the possible UV complete models that FASER as a LSW experiment will be able to explore, we find that the projections are not able to reach the KSVZ and DFSZ axion benchmarks and they barely touch the preferred QCD axion window~\cite{Georgi:1986df, DiLuzio:2016sbl, DiLuzio:2017pfr}, which is illustrated by the diagonal yellow shaded region. Therefore this proposal mainly probes other axion-like particles. Nonetheless, FASER could still probe QCD axions that solve the strong CP problem in the context of photophilic axion models where the axion coupling to photon is enhanced~\cite{Farina:2016tgd, Darme:2020gyx, Agrawal:2021dbo} or the $Z_{\mathcal{N}}$ axion model~\cite{Hook:2018jle, DiLuzio:2021pxd, DiLuzio:2021gos} where the axion mass is suppressed. \medskip 

Another interesting feature of our proposal is the ability of measuring the $\axion$ mass given a positive signal. Both helioscopes and LSW experiments, provide the best bounds in the maximum coherence limit $qL\ll 1$ and when that condition is fulfilled they are not sensitive to the $\axion$ mass. 
However, for $\axion$ masses such that $qL\sim 1$ the conversion probability depends on the energy of the photon (as one can see by substituting $q\simeq m_a^2/(2E_\gamma)$ in \cref{Eq:axionphoton conversion probability}). This oscillatory behaviour as a function of the photon energy depends on the axion mass and thus may allow to extract it~\cite{Dafni:2018tvj, Jaeckel:2018mbn}. 
Our proposal is most competitive in the region where the $\axion$-photon conversion is suppressed due to the loss of coherence $qL\gtrsim \mathcal{O}(1)$ and therefore in the hypothetical scenario of a positive detection we could benefit from the excellent energy resolution of the FASER calorimeter~\cite{FASER:2018bac} of about $1\%$ to extract the $\axion$ mass.

It is also worth emphasizing that the coherent $\axion$-photon oscillation in a magnetic field could be relevant for other $\axion$ searches where the $\axion$ couples not only to photons but also to other particles such as electrons or heavy electroweak gauge bosons. For example, some searches could use coherent conversion for $\axion$ production and $\axion$ decay or scattering via the coupling to electrons for detection. Alternatively, $\axions$ could be produced through another mechanism (e.g. rare meson decay via the $\axion$ coupling to electroweak gauge bosons or quarks) and be detected via coherent conversion~\cite{Gavela:2019wzg,Gori:2020xvq, Kling:2020mch}.

Regarding the outlook and possible improvements of this idea, there exist several techniques that could enhance the reach of this proposal and have already been proposed in the past for helioscopes or LSW~\cite{Arias:2010bh, CAST:2013bqn}. One option is the use of a buffer gas in the conversion cavity that generates an effective photon mass that maintains the coherence of the oscillation for larger masses. Another option is the use of non-homogeneous magnetic field \cite{Arias:2010bh, Fukuda:2022not} or the use of resonant cavities in the conversion chambers~\cite{Redondo:2010dp, ALPS:2009des, Ehret:2010mh}. Although it does not seem trivial to implement the buffer gas and the resonant cavity  improvements for such energetic photons, the use of non-homogeneous magnetic field profiles might be a viable option at the FPF.

\vspace{0.2cm}
\begin{acknowledgments}
\vspace{-0.4cm}
We thank Axel Lindner for discussions on the strategies to improve the sensitivity of LSW experiments. We also thank Javier Redondo and Igor Irastorza for useful comments regarding the endpoint of the CAST bound.  We thank Andrea Caputo for clarifications regarding the extent of the Supernova bound. F.K. and P.Q.~acknowledge support by the Deutsche Forschungsgemeinschaft under Germany's Excellence Strategy - EXC 2121 Quantum Universe - 390833306. This project has received funding/support from the European Union’s Horizon 2020 research and innovation programme under the Marie Sklodowska-Curie grant agreement No 860881-HIDDeN.
\end{acknowledgments}

\bibliographystyle{utphys}
\bibliography{bibliography}

\providecommand{\href}[2]{#2}\begingroup\raggedright\begin{thebibliography}{100}

\bibitem{Peccei:1977hh}
R.~D. Peccei and H.~R. Quinn, ``{CP Conservation in the Presence of
  Instantons},'' \href{http://dx.doi.org/10.1103/PhysRevLett.38.1440}{{\em
  Phys. Rev. Lett.} {\bfseries 38} (1977) 1440--1443}.
[,328(1977)].

\bibitem{Kim:1979if}
J.~E. Kim, ``{Weak Interaction Singlet and Strong CP Invariance},''
  \href{http://dx.doi.org/10.1103/PhysRevLett.43.103}{{\em Phys. Rev. Lett.}
  {\bfseries 43} (1979) 103}.

\bibitem{Shifman:1979if}
M.~A. Shifman, A.~I. Vainshtein, and V.~I. Zakharov, ``{Can Confinement Ensure
  Natural CP Invariance of Strong Interactions?},''
\href{http://dx.doi.org/10.1016/0550-3213(80)90209-6}{{\em Nucl. Phys.}
  {\bfseries B166} (1980) 493--506}.

\bibitem{Zhitnitsky:1980tq}
A.~R. Zhitnitsky, ``{On Possible Suppression of the Axion Hadron Interactions.
  (In Russian)},'' {\em Sov. J. Nucl. Phys.} {\bfseries 31} (1980) 260.
[Yad. Fiz.31,497(1980)].

\bibitem{Dine:1981rt}
M.~Dine, W.~Fischler, and M.~Srednicki, ``{A Simple Solution to the Strong CP
  Problem with a Harmless Axion},''
\href{http://dx.doi.org/10.1016/0370-2693(81)90590-6}{{\em Phys. Lett.}
  {\bfseries B104} (1981) 199--202}.

\bibitem{Kim:1984pt}
J.~E. Kim, ``{A Composite Invisible Axion},''
\href{http://dx.doi.org/10.1103/PhysRevD.31.1733}{{\em Phys. Rev.} {\bfseries
  D31} (1985) 1733}.

\bibitem{Dienes:1999gw}
K.~R. Dienes, E.~Dudas, and T.~Gherghetta, ``{Invisible axions and large radius
  compactifications},''
  \href{http://dx.doi.org/10.1103/PhysRevD.62.105023}{{\em Phys. Rev. D}
  {\bfseries 62} (2000) 105023},
  \href{http://arxiv.org/abs/hep-ph/9912455}{{\ttfamily arXiv:hep-ph/9912455}}.

\bibitem{Gelmini:1980re}
G.~B. Gelmini and M.~Roncadelli, ``{Left-Handed Neutrino Mass Scale and
  Spontaneously Broken Lepton Number},''
\href{http://dx.doi.org/10.1016/0370-2693(81)90559-1}{{\em Phys. Lett.}
  {\bfseries 99B} (1981) 411--415}.

\bibitem{Davidson:1981zd}
A.~Davidson and K.~C. Wali, ``{Minimal flavour unification via
  multigenerational Peccei-Quinn symmetry},''
\href{http://dx.doi.org/10.1103/PhysRevLett.48.11}{{\em Phys. Rev. Lett.}
  {\bfseries 48} (1982) 11}.

\bibitem{Wilczek:1982rv}
F.~Wilczek, ``{Axions and Family Symmetry Breaking},''
\href{http://dx.doi.org/10.1103/PhysRevLett.49.1549}{{\em Phys. Rev. Lett.}
  {\bfseries 49} (1982) 1549--1552}.

\bibitem{Cicoli:2013ana}
M.~Cicoli,
  \href{http://dx.doi.org/10.3204/DESY-PROC-2013-04/cicoli_michele}{``{Axion-like
  Particles from String Compactifications},''} in {\em {Proceedings, 9th Patras
  Workshop on Axions, WIMPs and WISPs (AXION-WIMP 2013): Mainz, Germany, June
  24-28, 2013}}, pp.~235--242.
\newblock 2013.
\newblock
\href{http://arxiv.org/abs/1309.6988}{{\ttfamily arXiv:1309.6988 [hep-th]}}.
\newblock

\bibitem{Abbott:1982af}
L.~F. Abbott and P.~Sikivie, ``{A Cosmological Bound on the Invisible Axion},''
  \href{http://dx.doi.org/10.1016/0370-2693(83)90638-X}{{\em Phys. Lett.}
  {\bfseries B120} (1983) 133--136}.
[,URL(1982)].

\bibitem{Dine:1982ah}
M.~Dine and W.~Fischler, ``{The Not So Harmless Axion},''
  \href{http://dx.doi.org/10.1016/0370-2693(83)90639-1}{{\em Phys. Lett. B}
  {\bfseries 120} (1983) 137--141}.

\bibitem{Preskill:1982cy}
J.~Preskill, M.~B. Wise, and F.~Wilczek, ``{Cosmology of the Invisible
  Axion},'' \href{http://dx.doi.org/10.1016/0370-2693(83)90637-8}{{\em Phys.
  Lett. B} {\bfseries 120} (1983) 127--132}.

\bibitem{Spector:2019ooq}
{\bfseries ALPS} Collaboration, A.~Spector, ``{ALPSII Status Report},'' in {\em
  {14th Patras Workshop on Axions, WIMPs and WISPs}}.
\newblock 6, 2019.
\newblock \href{http://arxiv.org/abs/1906.09011}{{\ttfamily arXiv:1906.09011
  [physics.ins-det]}}.

\bibitem{Irastorza:2013dav}
{\bfseries IAXO} Collaboration, I.~Irastorza {\em et~al.}, ``{The International
  Axion Observatory IAXO. Letter of Intent to the CERN SPS committee},''.

\bibitem{Anastassopoulos:2017ftl}
{\bfseries CAST} Collaboration, V.~Anastassopoulos {\em et~al.}, ``{New CAST
  Limit on the Axion-Photon Interaction},''
  \href{http://dx.doi.org/10.1038/nphys4109}{{\em Nature Phys.} {\bfseries 13}
  (2017) 584--590}, \href{http://arxiv.org/abs/1705.02290}{{\ttfamily
  arXiv:1705.02290 [hep-ex]}}.

\bibitem{Armengaud:2014gea}
E.~Armengaud {\em et~al.}, ``{Conceptual Design of the International Axion
  Observatory (IAXO)},''
  \href{http://dx.doi.org/10.1088/1748-0221/9/05/T05002}{{\em JINST} {\bfseries
  9} (2014) T05002},
\href{http://arxiv.org/abs/1401.3233}{{\ttfamily arXiv:1401.3233
  [physics.ins-det]}}.

\bibitem{Giannotti:2016drd}
M.~Giannotti, J.~Ruz, and J.~K. Vogel, ``{IAXO, next-generation of
  helioscopes},'' \href{http://dx.doi.org/10.22323/1.282.0195}{{\em PoS}
  {\bfseries ICHEP2016} (2016) 195},
\href{http://arxiv.org/abs/1611.04652}{{\ttfamily arXiv:1611.04652
  [physics.ins-det]}}.

\bibitem{Majorovits:2017ppy}
{\bfseries MADMAX interest Group} Collaboration, B.~Majorovits {\em et~al.},
  ``{MADMAX: A new road to axion dark matter detection},'' in {\em {15th
  International Conference on Topics in Astroparticle and Underground Physics
  (TAUP 2017) Sudbury, Ontario, Canada, July 24-28, 2017}}.
\newblock 2017.
\newblock
\href{http://arxiv.org/abs/1712.01062}{{\ttfamily arXiv:1712.01062
  [physics.ins-det]}}.
\newblock

\bibitem{Budker:2013hfa}
D.~Budker, P.~W. Graham, M.~Ledbetter, S.~Rajendran, and A.~Sushkov,
  ``{Proposal for a Cosmic Axion Spin Precession Experiment (CASPEr)},''
  \href{http://dx.doi.org/10.1103/PhysRevX.4.021030}{{\em Phys. Rev.}
  {\bfseries X4} no.~2, (2014) 021030},
\href{http://arxiv.org/abs/1306.6089}{{\ttfamily arXiv:1306.6089 [hep-ph]}}.

\bibitem{Ruoso:2015ytk}
G.~Ruoso, A.~Lombardi, A.~Ortolan, R.~Pengo, C.~Braggio, G.~Carugno, C.~S.
  Gallo, and C.~C. Speake, ``{The QUAX proposal: a search of galactic axion
  with magnetic materials},''
  \href{http://dx.doi.org/10.1088/1742-6596/718/4/042051}{{\em J. Phys. Conf.
  Ser.} {\bfseries 718} no.~4, (2016) 042051},
\href{http://arxiv.org/abs/1511.09461}{{\ttfamily arXiv:1511.09461 [hep-ph]}}.

\bibitem{Experiment:2017icw}
{\bfseries FUNK Experiment} Collaboration, D.~Veberič {\em et~al.}, ``{Search
  for hidden-photon dark matter with the FUNK experiment},''
  \href{http://dx.doi.org/10.22323/1.301.0880}{{\em PoS} {\bfseries ICRC2017}
  (2018) 880},
\href{http://arxiv.org/abs/1711.02958}{{\ttfamily arXiv:1711.02958 [hep-ex]}}.

\bibitem{Zenner:2013rta}
{\bfseries nEDM} Collaboration, J.~Zenner, ``{The nEDM experiment at the Paul
  Scherrer Institute, Switzerland},''
\href{http://dx.doi.org/10.1063/1.4826767}{{\em AIP Conf. Proc.} {\bfseries
  1560} no.~1, (2013) 254--256}.

\bibitem{Adams:2022pbo}
C.~B. Adams {\em et~al.}, ``{Axion Dark Matter},''
  \href{http://arxiv.org/abs/2203.14923}{{\ttfamily arXiv:2203.14923
  [hep-ex]}}.

\bibitem{Bauer:2017ris}
M.~Bauer, M.~Neubert, and A.~Thamm, ``{Collider Probes of Axion-Like
  Particles},'' \href{http://dx.doi.org/10.1007/JHEP12(2017)044}{{\em JHEP}
  {\bfseries 12} (2017) 044},
\href{http://arxiv.org/abs/1708.00443}{{\ttfamily arXiv:1708.00443 [hep-ph]}}.

\bibitem{Alonso-Alvarez:2018irt}
G.~Alonso-\'Alvarez, M.~B. Gavela, and P.~Quilez, ``{Axion couplings to
  electroweak gauge bosons},''
  \href{http://dx.doi.org/10.1140/epjc/s10052-019-6732-5}{{\em Eur. Phys. J. C}
  {\bfseries 79} no.~3, (2019) 223},
\href{http://arxiv.org/abs/1811.05466}{{\ttfamily arXiv:1811.05466 [hep-ph]}}.

\bibitem{Izaguirre:2016dfi}
E.~Izaguirre, T.~Lin, and B.~Shuve, ``{Searching for Axionlike Particles in
  Flavor-Changing Neutral Current Processes},''
  \href{http://dx.doi.org/10.1103/PhysRevLett.118.111802}{{\em Phys. Rev.
  Lett.} {\bfseries 118} no.~11, (2017) 111802},
\href{http://arxiv.org/abs/1611.09355}{{\ttfamily arXiv:1611.09355 [hep-ph]}}.

\bibitem{CidVidal:2018blh}
X.~Cid~Vidal, A.~Mariotti, D.~Redigolo, F.~Sala, and K.~Tobioka, ``{New Axion
  Searches at Flavor Factories},''
  \href{http://dx.doi.org/10.1007/JHEP01(2019)113}{{\em JHEP} {\bfseries 01}
  (2019) 113}, \href{http://arxiv.org/abs/1810.09452}{{\ttfamily
  arXiv:1810.09452 [hep-ph]}}.
[Erratum: JHEP 06, 141 (2020)].

\bibitem{Gavela:2019wzg}
M.~B. Gavela, R.~Houtz, P.~Quilez, R.~Del~Rey, and O.~Sumensari, ``{Flavor
  constraints on electroweak ALP couplings},''
  \href{http://dx.doi.org/10.1140/epjc/s10052-019-6889-y}{{\em Eur. Phys. J. C}
  {\bfseries 79} no.~5, (2019) 369},
\href{http://arxiv.org/abs/1901.02031}{{\ttfamily arXiv:1901.02031 [hep-ph]}}.

\bibitem{Bauer:2021mvw}
M.~Bauer, M.~Neubert, S.~Renner, M.~Schnubel, and A.~Thamm, ``{Flavor probes of
  axion-like particles},'' \href{http://arxiv.org/abs/2110.10698}{{\ttfamily
  arXiv:2110.10698 [hep-ph]}}.

\bibitem{Dobrich:2019dxc}
B.~D\"obrich, J.~Jaeckel, and T.~Spadaro, ``{Light in the beam dump - ALP
  production from decay photons in proton beam-dumps},''
  \href{http://dx.doi.org/10.1007/JHEP05(2019)213}{{\em JHEP} {\bfseries 05}
  (2019) 213}, \href{http://arxiv.org/abs/1904.02091}{{\ttfamily
  arXiv:1904.02091 [hep-ph]}}. [Erratum: JHEP 10, 046 (2020)].

\bibitem{Brdar:2020dpr}
V.~Brdar, B.~Dutta, W.~Jang, D.~Kim, I.~M. Shoemaker, Z.~Tabrizi, A.~Thompson,
  and J.~Yu, ``{Axionlike Particles at Future Neutrino Experiments: Closing the
  Cosmological Triangle},''
  \href{http://dx.doi.org/10.1103/PhysRevLett.126.201801}{{\em Phys. Rev.
  Lett.} {\bfseries 126} no.~20, (2021) 201801},
  \href{http://arxiv.org/abs/2011.07054}{{\ttfamily arXiv:2011.07054
  [hep-ph]}}.

\bibitem{Coloma:2022hlv}
P.~Coloma, P.~Hern\'andez, and S.~Urrea, ``{New bounds on axion-like particles
  from MicroBooNE},'' \href{http://arxiv.org/abs/2202.03447}{{\ttfamily
  arXiv:2202.03447 [hep-ph]}}.

\bibitem{Raffelt:1990yz}
G.~G. Raffelt, ``{Astrophysical methods to constrain axions and other novel
  particle phenomena},''
  \href{http://dx.doi.org/10.1016/0370-1573(90)90054-6}{{\em Phys. Rept.}
  {\bfseries 198} (1990) 1--113}.

\bibitem{Bar:2019ifz}
N.~Bar, K.~Blum, and G.~D'Amico, ``{Is there a supernova bound on axions?},''
  \href{http://dx.doi.org/10.1103/PhysRevD.101.123025}{{\em Phys. Rev. D}
  {\bfseries 101} no.~12, (2020) 123025},
\href{http://arxiv.org/abs/1907.05020}{{\ttfamily arXiv:1907.05020 [hep-ph]}}.

\bibitem{Caputo:2021rux}
A.~Caputo, G.~Raffelt, and E.~Vitagliano, ``{Muonic boson limits: Supernova
  redux},'' \href{http://dx.doi.org/10.1103/PhysRevD.105.035022}{{\em Phys.
  Rev. D} {\bfseries 105} no.~3, (2022) 035022},
  \href{http://arxiv.org/abs/2109.03244}{{\ttfamily arXiv:2109.03244
  [hep-ph]}}.

\bibitem{Depta:2020wmr}
P.~F. Depta, M.~Hufnagel, and K.~Schmidt-Hoberg, ``{Robust cosmological
  constraints on axion-like particles},''
  \href{http://dx.doi.org/10.1088/1475-7516/2020/05/009}{{\em JCAP} {\bfseries
  05} (2020) 009}, \href{http://arxiv.org/abs/2002.08370}{{\ttfamily
  arXiv:2002.08370 [hep-ph]}}.

\bibitem{Carenza:2020zil}
P.~Carenza, O.~Straniero, B.~D\"obrich, M.~Giannotti, G.~Lucente, and
  A.~Mirizzi, ``{Constraints on the coupling with photons of heavy
  axion-like-particles from Globular Clusters},''
  \href{http://dx.doi.org/10.1016/j.physletb.2020.135709}{{\em Phys. Lett. B}
  {\bfseries 809} (2020) 135709},
  \href{http://arxiv.org/abs/2004.08399}{{\ttfamily arXiv:2004.08399
  [hep-ph]}}.

\bibitem{Jaeckel:2006xm}
J.~Jaeckel, E.~Masso, J.~Redondo, A.~Ringwald, and F.~Takahashi, ``{The Need
  for purely laboratory-based axion-like particle searches},''
  \href{http://dx.doi.org/10.1103/PhysRevD.75.013004}{{\em Phys. Rev. D}
  {\bfseries 75} (2007) 013004},
  \href{http://arxiv.org/abs/hep-ph/0610203}{{\ttfamily arXiv:hep-ph/0610203}}.

\bibitem{Feng:2017uoz}
J.~L. Feng, I.~Galon, F.~Kling, and S.~Trojanowski, ``{ForwArd Search
  ExpeRiment at the LHC},''
  \href{http://dx.doi.org/10.1103/PhysRevD.97.035001}{{\em Phys. Rev. D}
  {\bfseries 97} no.~3, (2018) 035001},
  \href{http://arxiv.org/abs/1708.09389}{{\ttfamily arXiv:1708.09389
  [hep-ph]}}.

\bibitem{FASER:2018ceo}
{\bfseries FASER} Collaboration, A.~Ariga {\em et~al.}, ``{Letter of Intent for
  FASER: ForwArd Search ExpeRiment at the LHC},''
  \href{http://arxiv.org/abs/1811.10243}{{\ttfamily arXiv:1811.10243
  [physics.ins-det]}}.

\bibitem{FASER:2018bac}
{\bfseries FASER} Collaboration, A.~Ariga {\em et~al.}, ``{Technical Proposal
  for FASER: ForwArd Search ExpeRiment at the LHC},''
  \href{http://arxiv.org/abs/1812.09139}{{\ttfamily arXiv:1812.09139
  [physics.ins-det]}}.

\bibitem{Kling:2021gos}
F.~Kling and L.~J. Nevay, ``{Forward neutrino fluxes at the LHC},''
  \href{http://dx.doi.org/10.1103/PhysRevD.104.113008}{{\em Phys. Rev. D}
  {\bfseries 104} no.~11, (2021) 113008},
  \href{http://arxiv.org/abs/2105.08270}{{\ttfamily arXiv:2105.08270
  [hep-ph]}}.

\bibitem{ZurbanoFernandez:2020cco}
I.~Zurbano~Fernandez {\em et~al.}, ``{High-Luminosity Large Hadron Collider
  (HL-LHC): Technical design report},''.

\bibitem{Anchordoqui:2021ghd}
L.~A. Anchordoqui {\em et~al.}, ``{The Forward Physics Facility: Sites,
  Experiments, and Physics Potential},''
  \href{http://arxiv.org/abs/2109.10905}{{\ttfamily arXiv:2109.10905
  [hep-ph]}}.

\bibitem{Feng:2022inv}
J.~L. Feng {\em et~al.}, ``{The Forward Physics Facility at the High-Luminosity
  LHC},'' in {\em {2022 Snowmass Summer Study}}.
\newblock 3, 2022.
\newblock \href{http://arxiv.org/abs/2203.05090}{{\ttfamily arXiv:2203.05090
  [hep-ex]}}.

\bibitem{ATLAS:2012yve}
{\bfseries ATLAS} Collaboration, G.~Aad {\em et~al.}, ``{Observation of a new
  particle in the search for the Standard Model Higgs boson with the ATLAS
  detector at the LHC},''
  \href{http://dx.doi.org/10.1016/j.physletb.2012.08.020}{{\em Phys. Lett. B}
  {\bfseries 716} (2012) 1--29},
  \href{http://arxiv.org/abs/1207.7214}{{\ttfamily arXiv:1207.7214 [hep-ex]}}.

\bibitem{CMS:2012qbp}
{\bfseries CMS} Collaboration, S.~Chatrchyan {\em et~al.}, ``{Observation of a
  New Boson at a Mass of 125 GeV with the CMS Experiment at the LHC},''
  \href{http://dx.doi.org/10.1016/j.physletb.2012.08.021}{{\em Phys. Lett. B}
  {\bfseries 716} (2012) 30--61},
  \href{http://arxiv.org/abs/1207.7235}{{\ttfamily arXiv:1207.7235 [hep-ex]}}.

\bibitem{Kling:2021fwx}
F.~Kling and S.~Trojanowski, ``{Forward experiment sensitivity estimator for
  the LHC and future hadron colliders},''
  \href{http://dx.doi.org/10.1103/PhysRevD.104.035012}{{\em Phys. Rev. D}
  {\bfseries 104} no.~3, (2021) 035012},
  \href{http://arxiv.org/abs/2105.07077}{{\ttfamily arXiv:2105.07077
  [hep-ph]}}.

\bibitem{Pierog:2013ria}
T.~Pierog, I.~Karpenko, J.~M. Katzy, E.~Yatsenko, and K.~Werner, ``{EPOS LHC:
  Test of collective hadronization with data measured at the CERN Large Hadron
  Collider},'' \href{http://dx.doi.org/10.1103/PhysRevC.92.034906}{{\em Phys.
  Rev. C} {\bfseries 92} no.~3, (2015) 034906},
  \href{http://arxiv.org/abs/1306.0121}{{\ttfamily arXiv:1306.0121 [hep-ph]}}.

\bibitem{CRMC}
R.~Ulrich, T.~Pierog, and C.~Baus, ``Cosmic ray monte carlo package, crmc,''
  Aug., 2021.
\newblock \url{https://doi.org/10.5281/zenodo.5270381}.

\bibitem{Nevay:2018zhp}
L.~J. Nevay {\em et~al.}, ``{BDSIM: An accelerator tracking code with
  particle\textendash{}matter interactions},''
  \href{http://dx.doi.org/10.1016/j.cpc.2020.107200}{{\em Comput. Phys.
  Commun.} {\bfseries 252} (2020) 107200},
  \href{http://arxiv.org/abs/1808.10745}{{\ttfamily arXiv:1808.10745
  [physics.comp-ph]}}.

\bibitem{Redondo:2010dp}
J.~Redondo and A.~Ringwald, ``{Light shining through walls},''
  \href{http://dx.doi.org/10.1080/00107514.2011.563516}{{\em Contemp. Phys.}
  {\bfseries 52} (2011) 211--236},
  \href{http://arxiv.org/abs/1011.3741}{{\ttfamily arXiv:1011.3741 [hep-ph]}}.

\bibitem{Primakoff:1951iae}
H.~Primakoff, ``{Photoproduction of neutral mesons in nuclear electric fields
  and the mean life of the neutral meson},''
  \href{http://dx.doi.org/10.1103/PhysRev.81.899}{{\em Phys. Rev.} {\bfseries
  81} (1951) 899}.

\bibitem{Tsai:1986tx}
Y.-S. Tsai, ``{AXION BREMSSTRAHLUNG BY AN ELECTRON BEAM},''
  \href{http://dx.doi.org/10.1103/PhysRevD.34.1326}{{\em Phys. Rev. D}
  {\bfseries 34} (1986) 1326}.

\bibitem{Feng:2018pew}
J.~L. Feng, I.~Galon, F.~Kling, and S.~Trojanowski, ``{Axionlike particles at
  FASER: The LHC as a photon beam dump},''
  \href{http://dx.doi.org/10.1103/PhysRevD.98.055021}{{\em Phys. Rev. D}
  {\bfseries 98} no.~5, (2018) 055021},
  \href{http://arxiv.org/abs/1806.02348}{{\ttfamily arXiv:1806.02348
  [hep-ph]}}.

\bibitem{FASER:2019aik}
{\bfseries FASER} Collaboration, A.~Ariga {\em et~al.}, ``{FASER: ForwArd
  Search ExpeRiment at the LHC},''
  \href{http://arxiv.org/abs/1901.04468}{{\ttfamily arXiv:1901.04468
  [hep-ex]}}.

\bibitem{FASER:2021cpr}
{\bfseries FASER} Collaboration, H.~Abreu {\em et~al.}, ``{The trigger and data
  acquisition system of the FASER experiment},''
  \href{http://dx.doi.org/10.1088/1748-0221/16/12/P12028}{{\em JINST}
  {\bfseries 16} no.~12, (2021) P12028},
  \href{http://arxiv.org/abs/2110.15186}{{\ttfamily arXiv:2110.15186
  [physics.ins-det]}}.

\bibitem{FASER:2021ljd}
{\bfseries FASER} Collaboration, H.~Abreu {\em et~al.}, ``{The tracking
  detector of the FASER experiment},''
  \href{http://arxiv.org/abs/2112.01116}{{\ttfamily arXiv:2112.01116
  [physics.ins-det]}}.

\bibitem{Feng:2017vli}
J.~L. Feng, I.~Galon, F.~Kling, and S.~Trojanowski, ``{Dark Higgs bosons at the
  ForwArd Search ExpeRiment},''
  \href{http://dx.doi.org/10.1103/PhysRevD.97.055034}{{\em Phys. Rev. D}
  {\bfseries 97} no.~5, (2018) 055034},
  \href{http://arxiv.org/abs/1710.09387}{{\ttfamily arXiv:1710.09387
  [hep-ph]}}.

\bibitem{Kling:2018wct}
F.~Kling and S.~Trojanowski, ``{Heavy Neutral Leptons at FASER},''
  \href{http://dx.doi.org/10.1103/PhysRevD.97.095016}{{\em Phys. Rev. D}
  {\bfseries 97} no.~9, (2018) 095016},
  \href{http://arxiv.org/abs/1801.08947}{{\ttfamily arXiv:1801.08947
  [hep-ph]}}.

\bibitem{Feng:2018noy}
J.~L. Feng, I.~Galon, F.~Kling, and S.~Trojanowski, ``{Axionlike particles at
  FASER: The LHC as a photon beam dump},''
  \href{http://dx.doi.org/10.1103/PhysRevD.98.055021}{{\em Phys. Rev.}
  {\bfseries D98} no.~5, (2018) 055021},
\href{http://arxiv.org/abs/1806.02348}{{\ttfamily arXiv:1806.02348 [hep-ph]}}.

\bibitem{FASER:2018eoc}
{\bfseries FASER} Collaboration, A.~Ariga {\em et~al.},
  ``{FASER\textquoteright{}s physics reach for long-lived particles},''
  \href{http://dx.doi.org/10.1103/PhysRevD.99.095011}{{\em Phys. Rev. D}
  {\bfseries 99} no.~9, (2019) 095011},
  \href{http://arxiv.org/abs/1811.12522}{{\ttfamily arXiv:1811.12522
  [hep-ph]}}.

\bibitem{FASER:2019dxq}
{\bfseries FASER} Collaboration, H.~Abreu {\em et~al.}, ``{Detecting and
  Studying High-Energy Collider Neutrinos with FASER at the LHC},''
  \href{http://dx.doi.org/10.1140/epjc/s10052-020-7631-5}{{\em Eur. Phys. J. C}
  {\bfseries 80} no.~1, (2020) 61},
  \href{http://arxiv.org/abs/1908.02310}{{\ttfamily arXiv:1908.02310
  [hep-ex]}}.

\bibitem{FASER:2020gpr}
{\bfseries FASER} Collaboration, H.~Abreu {\em et~al.}, ``{Technical Proposal:
  FASERnu},'' \href{http://arxiv.org/abs/2001.03073}{{\ttfamily
  arXiv:2001.03073 [physics.ins-det]}}.

\bibitem{FASER:2021mtu}
{\bfseries FASER} Collaboration, H.~Abreu {\em et~al.}, ``{First neutrino
  interaction candidates at the LHC},''
  \href{http://dx.doi.org/10.1103/PhysRevD.104.L091101}{{\em Phys. Rev. D}
  {\bfseries 104} no.~9, (2021) L091101},
  \href{http://arxiv.org/abs/2105.06197}{{\ttfamily arXiv:2105.06197
  [hep-ex]}}.

\bibitem{Boyd:2803084}
J.~Boyd, ``{The FASER W-Si High Precision Preshower Technical Proposal},''
  tech. rep., CERN, Geneva, Mar, 2022.
\newblock \url{https://cds.cern.ch/record/2803084}.

\bibitem{GrillidiCortona:2015jxo}
G.~Grilli~di Cortona, E.~Hardy, J.~Pardo~Vega, and G.~Villadoro, ``{The QCD
  axion, precisely},'' \href{http://dx.doi.org/10.1007/JHEP01(2016)034}{{\em
  JHEP} {\bfseries 01} (2016) 034},
  \href{http://arxiv.org/abs/1511.02867}{{\ttfamily arXiv:1511.02867
  [hep-ph]}}.

\bibitem{AxionLimits}
C.~O'Hare, ``cajohare/axionlimits: Axionlimits,''.
  \url{https://doi.org/10.5281/zenodo.3932430}.

\bibitem{Jodlowski:2020vhr}
K.~Jod\l{}owski and S.~Trojanowski, ``{Neutrino beam-dump experiment with FASER
  at the LHC},'' \href{http://dx.doi.org/10.1007/JHEP05(2021)191}{{\em JHEP}
  {\bfseries 05} (2021) 191}, \href{http://arxiv.org/abs/2011.04751}{{\ttfamily
  arXiv:2011.04751 [hep-ph]}}.

\bibitem{CAST:2007jps}
{\bfseries CAST} Collaboration, S.~Andriamonje {\em et~al.}, ``{An Improved
  limit on the axion-photon coupling from the CAST experiment},''
  \href{http://dx.doi.org/10.1088/1475-7516/2007/04/010}{{\em JCAP} {\bfseries
  04} (2007) 010}, \href{http://arxiv.org/abs/hep-ex/0702006}{{\ttfamily
  arXiv:hep-ex/0702006}}.

\bibitem{CAST:2017uph}
{\bfseries CAST} Collaboration, V.~Anastassopoulos {\em et~al.}, ``{New CAST
  Limit on the Axion-Photon Interaction},''
  \href{http://dx.doi.org/10.1038/nphys4109}{{\em Nature Phys.} {\bfseries 13}
  (2017) 584--590}, \href{http://arxiv.org/abs/1705.02290}{{\ttfamily
  arXiv:1705.02290 [hep-ex]}}.

\bibitem{Shilon:2013xma}
I.~Shilon, A.~Dudarev, H.~Silva, U.~Wagner, and H.~H.~J. ten Kate, ``{The
  Superconducting Toroid for the New International AXion Observatory (IAXO)},''
  \href{http://dx.doi.org/10.1109/TASC.2013.2280654}{{\em IEEE Trans. Appl.
  Supercond.} {\bfseries 24} no.~3, (2014) 4500104},
  \href{http://arxiv.org/abs/1309.2117}{{\ttfamily arXiv:1309.2117
  [physics.ins-det]}}.

\bibitem{Ge:2020zww}
S.-F. Ge, K.~Hamaguchi, K.~Ichimura, K.~Ishidoshiro, Y.~Kanazawa, Y.~Kishimoto,
  N.~Nagata, and J.~Zheng, ``{Supernova-scope for the Direct Search of
  Supernova Axions},''
  \href{http://dx.doi.org/10.1088/1475-7516/2020/11/059}{{\em JCAP} {\bfseries
  11} (2020) 059}, \href{http://arxiv.org/abs/2008.03924}{{\ttfamily
  arXiv:2008.03924 [hep-ph]}}.

\bibitem{Ouellet:2018beu}
J.~L. Ouellet {\em et~al.}, ``{First Results from ABRACADABRA-10 cm: A Search
  for Sub-$\mu$eV Axion Dark Matter},''
  \href{http://dx.doi.org/10.1103/PhysRevLett.122.121802}{{\em Phys. Rev.
  Lett.} {\bfseries 122} no.~12, (2019) 121802},
  \href{http://arxiv.org/abs/1810.12257}{{\ttfamily arXiv:1810.12257
  [hep-ex]}}.

\bibitem{Salemi:2021gck}
C.~P. Salemi {\em et~al.}, ``{Search for Low-Mass Axion Dark Matter with
  ABRACADABRA-10~cm},''
  \href{http://dx.doi.org/10.1103/PhysRevLett.127.081801}{{\em Phys. Rev.
  Lett.} {\bfseries 127} no.~8, (2021) 081801},
  \href{http://arxiv.org/abs/2102.06722}{{\ttfamily arXiv:2102.06722
  [hep-ex]}}.

\bibitem{Asztalos2010}
S.~J. {Asztalos}, G.~{Carosi}, C.~{Hagmann}, D.~{Kinion}, K.~{van Bibber},
  M.~{Hotz}, L.~J. {Rosenberg}, G.~{Rybka}, J.~{Hoskins}, J.~{Hwang},
  P.~{Sikivie}, D.~B. {Tanner}, R.~{Bradley}, J.~{Clarke}, and {ADMX
  Collaboration}, ``{SQUID-Based Microwave Cavity Search for Dark-Matter
  Axions},'' \href{http://dx.doi.org/10.1103/PhysRevLett.104.041301}{{\em Phys.
  Rev. Lett.} {\bfseries 104} no.~4, (Jan., 2010) 041301},
  \href{http://arxiv.org/abs/0910.5914}{{\ttfamily arXiv:0910.5914
  [astro-ph.CO]}}.

\bibitem{ADMX:2018gho}
{\bfseries ADMX} Collaboration, N.~Du {\em et~al.}, ``{A Search for Invisible
  Axion Dark Matter with the Axion Dark Matter Experiment},''
  \href{http://dx.doi.org/10.1103/PhysRevLett.120.151301}{{\em Phys. Rev.
  Lett.} {\bfseries 120} no.~15, (2018) 151301},
  \href{http://arxiv.org/abs/1804.05750}{{\ttfamily arXiv:1804.05750
  [hep-ex]}}.

\bibitem{ADMX:2019uok}
{\bfseries ADMX} Collaboration, T.~Braine {\em et~al.}, ``{Extended Search for
  the Invisible Axion with the Axion Dark Matter Experiment},''
  \href{http://dx.doi.org/10.1103/PhysRevLett.124.101303}{{\em Phys. Rev.
  Lett.} {\bfseries 124} no.~10, (2020) 101303},
  \href{http://arxiv.org/abs/1910.08638}{{\ttfamily arXiv:1910.08638
  [hep-ex]}}.

\bibitem{ADMX:2021nhd}
{\bfseries ADMX} Collaboration, C.~Bartram {\em et~al.}, ``{Search for
  Invisible Axion Dark Matter in the 3.3\textendash{}4.2\,\,\ensuremath{\mu}eV
  Mass Range},'' \href{http://dx.doi.org/10.1103/PhysRevLett.127.261803}{{\em
  Phys. Rev. Lett.} {\bfseries 127} no.~26, (2021) 261803},
  \href{http://arxiv.org/abs/2110.06096}{{\ttfamily arXiv:2110.06096
  [hep-ex]}}.

\bibitem{ADMX:2018ogs}
{\bfseries ADMX} Collaboration, C.~Boutan {\em et~al.}, ``{Piezoelectrically
  Tuned Multimode Cavity Search for Axion Dark Matter},''
  \href{http://dx.doi.org/10.1103/PhysRevLett.121.261302}{{\em Phys. Rev.
  Lett.} {\bfseries 121} no.~26, (2018) 261302},
  \href{http://arxiv.org/abs/1901.00920}{{\ttfamily arXiv:1901.00920
  [hep-ex]}}.

\bibitem{Bartram:2021ysp}
C.~Bartram {\em et~al.}, ``{Dark Matter Axion Search Using a Josephson
  Traveling Wave Parametric Amplifier},''
  \href{http://arxiv.org/abs/2110.10262}{{\ttfamily arXiv:2110.10262
  [hep-ex]}}.

\bibitem{Crisosto:2019fcj}
N.~Crisosto, P.~Sikivie, N.~S. Sullivan, D.~B. Tanner, J.~Yang, and G.~Rybka,
  ``{ADMX SLIC: Results from a Superconducting $LC$ Circuit Investigating Cold
  Axions},'' \href{http://dx.doi.org/10.1103/PhysRevLett.124.241101}{{\em Phys.
  Rev. Lett.} {\bfseries 124} no.~24, (2020) 241101},
  \href{http://arxiv.org/abs/1911.05772}{{\ttfamily arXiv:1911.05772
  [astro-ph.CO]}}.

\bibitem{Lee:2020cfj}
S.~Lee, S.~Ahn, J.~Choi, B.~R. Ko, and Y.~K. Semertzidis, ``{Axion Dark Matter
  Search around 6.7 $\mu$eV},''
  \href{http://dx.doi.org/10.1103/PhysRevLett.124.101802}{{\em Phys. Rev.
  Lett.} {\bfseries 124} no.~10, (2020) 101802},
  \href{http://arxiv.org/abs/2001.05102}{{\ttfamily arXiv:2001.05102
  [hep-ex]}}.

\bibitem{Jeong:2020cwz}
J.~Jeong, S.~Youn, S.~Bae, J.~Kim, T.~Seong, J.~E. Kim, and Y.~K. Semertzidis,
  ``{Search for Invisible Axion Dark Matter with a Multiple-Cell Haloscope},''
  \href{http://dx.doi.org/10.1103/PhysRevLett.125.221302}{{\em Phys. Rev.
  Lett.} {\bfseries 125} no.~22, (2020) 221302},
  \href{http://arxiv.org/abs/2008.10141}{{\ttfamily arXiv:2008.10141
  [hep-ex]}}.

\bibitem{CAPP:2020utb}
{\bfseries CAPP} Collaboration, O.~Kwon {\em et~al.}, ``{First Results from an
  Axion Haloscope at CAPP around 10.7 $\mu$eV},''
  \href{http://dx.doi.org/10.1103/PhysRevLett.126.191802}{{\em Phys. Rev.
  Lett.} {\bfseries 126} no.~19, (2021) 191802},
  \href{http://arxiv.org/abs/2012.10764}{{\ttfamily arXiv:2012.10764
  [hep-ex]}}.

\bibitem{Devlin:2021fpq}
J.~A. Devlin {\em et~al.}, ``{Constraints on the Coupling between Axionlike
  Dark Matter and Photons Using an Antiproton Superconducting Tuned Detection
  Circuit in a Cryogenic Penning Trap},''
  \href{http://dx.doi.org/10.1103/PhysRevLett.126.041301}{{\em Phys. Rev.
  Lett.} {\bfseries 126} no.~4, (2021) 041301},
  \href{http://arxiv.org/abs/2101.11290}{{\ttfamily arXiv:2101.11290
  [astro-ph.CO]}}.

\bibitem{Grenet:2021vbb}
T.~Grenet, R.~Ballou, Q.~Basto, K.~Martineau, P.~Perrier, P.~Pugnat,
  J.~Quevillon, N.~Roch, and C.~Smith, ``{The Grenoble Axion Haloscope platform
  (GrAHal): development plan and first results},''
  \href{http://arxiv.org/abs/2110.14406}{{\ttfamily arXiv:2110.14406
  [hep-ex]}}.

\bibitem{HAYSTAC:2018rwy}
{\bfseries HAYSTAC} Collaboration, L.~Zhong {\em et~al.}, ``{Results from phase
  1 of the HAYSTAC microwave cavity axion experiment},''
  \href{http://dx.doi.org/10.1103/PhysRevD.97.092001}{{\em Phys. Rev. D}
  {\bfseries 97} no.~9, (2018) 092001},
  \href{http://arxiv.org/abs/1803.03690}{{\ttfamily arXiv:1803.03690
  [hep-ex]}}.

\bibitem{HAYSTAC:2020kwv}
{\bfseries HAYSTAC} Collaboration, K.~M. Backes {\em et~al.}, ``{A
  quantum-enhanced search for dark matter axions},''
  \href{http://dx.doi.org/10.1038/s41586-021-03226-7}{{\em Nature} {\bfseries
  590} no.~7845, (2021) 238--242},
  \href{http://arxiv.org/abs/2008.01853}{{\ttfamily arXiv:2008.01853
  [quant-ph]}}.

\bibitem{McAllister:2017lkb}
B.~T. McAllister, G.~Flower, E.~N. Ivanov, M.~Goryachev, J.~Bourhill, and M.~E.
  Tobar, ``{The ORGAN Experiment: An axion haloscope above 15 GHz},''
  \href{http://dx.doi.org/10.1016/j.dark.2017.09.010}{{\em Phys. Dark Univ.}
  {\bfseries 18} (2017) 67--72},
  \href{http://arxiv.org/abs/1706.00209}{{\ttfamily arXiv:1706.00209
  [physics.ins-det]}}.

\bibitem{Alesini:2019ajt}
D.~Alesini {\em et~al.}, ``{Galactic axions search with a superconducting
  resonant cavity},'' \href{http://dx.doi.org/10.1103/PhysRevD.99.101101}{{\em
  Phys. Rev. D} {\bfseries 99} no.~10, (2019) 101101},
  \href{http://arxiv.org/abs/1903.06547}{{\ttfamily arXiv:1903.06547
  [physics.ins-det]}}.

\bibitem{Alesini:2020vny}
D.~Alesini {\em et~al.}, ``{Search for invisible axion dark matter of mass
  m$_a=43~\mu$eV with the QUAX--$a\gamma$ experiment},''
  \href{http://dx.doi.org/10.1103/PhysRevD.103.102004}{{\em Phys. Rev. D}
  {\bfseries 103} no.~10, (2021) 102004},
  \href{http://arxiv.org/abs/2012.09498}{{\ttfamily arXiv:2012.09498
  [hep-ex]}}.

\bibitem{CAST:2020rlf}
{\bfseries CAST} Collaboration, A.~A. Melc\'on {\em et~al.}, ``{First results
  of the CAST-RADES haloscope search for axions at 34.67 $\mu$eV},''
  \href{http://dx.doi.org/10.1007/JHEP10(2021)075}{{\em JHEP} {\bfseries 21}
  (2020) 075}, \href{http://arxiv.org/abs/2104.13798}{{\ttfamily
  arXiv:2104.13798 [hep-ex]}}.

\bibitem{DePanfilis}
S.~DePanfilis, A.~C. Melissinos, B.~E. Moskowitz, J.~T. Rogers, Y.~K.
  Semertzidis, W.~U. Wuensch, H.~J. Halama, A.~G. Prodell, W.~B. Fowler, and
  F.~A. Nezrick, ``Limits on the abundance and coupling of cosmic axions at
  4.5$<{m}_{a}<$5.0 \ensuremath{\mu}ev,''
  \href{http://dx.doi.org/10.1103/PhysRevLett.59.839}{{\em Phys. Rev. Lett.}
  {\bfseries 59} (Aug, 1987) 839--842}.
  \url{https://link.aps.org/doi/10.1103/PhysRevLett.59.839}.

\bibitem{Gramolin:2020ict}
A.~V. Gramolin, D.~Aybas, D.~Johnson, J.~Adam, and A.~O. Sushkov, ``{Search for
  axion-like dark matter with ferromagnets},''
  \href{http://dx.doi.org/10.1038/s41567-020-1006-6}{{\em Nature Phys.}
  {\bfseries 17} no.~1, (2021) 79--84},
  \href{http://arxiv.org/abs/2003.03348}{{\ttfamily arXiv:2003.03348
  [hep-ex]}}.

\bibitem{Arza:2021rrm}
A.~Arza, M.~A. Fedderke, P.~W. Graham, D.~F. Jackson~Kimball, and S.~Kalia,
  ``{Earth as a transducer for axion dark-matter detection},''
  \href{http://arxiv.org/abs/2112.09620}{{\ttfamily arXiv:2112.09620
  [hep-ph]}}.

\bibitem{Hagmann}
C.~Hagmann, P.~Sikivie, N.~S. Sullivan, and D.~B. Tanner, ``Results from a
  search for cosmic axions,''
  \href{http://dx.doi.org/10.1103/PhysRevD.42.1297}{{\em Phys. Rev. D}
  {\bfseries 42} (Aug, 1990) 1297--1300}.
  \url{https://link.aps.org/doi/10.1103/PhysRevD.42.1297}.

\bibitem{Thomson:2019aht}
C.~A. Thomson, B.~T. McAllister, M.~Goryachev, E.~N. Ivanov, and M.~E. Tobar,
  ``{Upconversion Loop Oscillator Axion Detection Experiment: A Precision
  Frequency Interferometric Axion Dark Matter Search with a Cylindrical
  Microwave Cavity},''
  \href{http://dx.doi.org/10.1103/PhysRevLett.127.019901}{{\em Phys. Rev.
  Lett.} {\bfseries 126} no.~8, (2021) 081803},
  \href{http://arxiv.org/abs/1912.07751}{{\ttfamily arXiv:1912.07751
  [hep-ex]}}. [Erratum: Phys.Rev.Lett. 127, 019901 (2021)].

\bibitem{Ehret:2010mh}
K.~Ehret {\em et~al.}, ``{New ALPS Results on Hidden-Sector Lightweights},''
  \href{http://dx.doi.org/10.1016/j.physletb.2010.04.066}{{\em Phys. Lett. B}
  {\bfseries 689} (2010) 149--155},
  \href{http://arxiv.org/abs/1004.1313}{{\ttfamily arXiv:1004.1313 [hep-ex]}}.

\bibitem{Betz:2013dza}
M.~Betz, F.~Caspers, M.~Gasior, M.~Thumm, and S.~W. Rieger, ``{First results of
  the CERN Resonant Weakly Interacting sub-eV Particle Search (CROWS)},''
  \href{http://dx.doi.org/10.1103/PhysRevD.88.075014}{{\em Phys. Rev. D}
  {\bfseries 88} no.~7, (2013) 075014},
  \href{http://arxiv.org/abs/1310.8098}{{\ttfamily arXiv:1310.8098
  [physics.ins-det]}}.

\bibitem{OSQAR:2015qdv}
{\bfseries OSQAR} Collaboration, R.~Ballou {\em et~al.}, ``{New exclusion
  limits on scalar and pseudoscalar axionlike particles from light shining
  through a wall},'' \href{http://dx.doi.org/10.1103/PhysRevD.92.092002}{{\em
  Phys. Rev. D} {\bfseries 92} no.~9, (2015) 092002},
  \href{http://arxiv.org/abs/1506.08082}{{\ttfamily arXiv:1506.08082
  [hep-ex]}}.

\bibitem{DellaValle:2015xxa}
F.~Della~Valle, A.~Ejlli, U.~Gastaldi, G.~Messineo, E.~Milotti, R.~Pengo,
  G.~Ruoso, and G.~Zavattini, ``{The PVLAS experiment: measuring vacuum
  magnetic birefringence and dichroism with a birefringent
  Fabry\textendash{}Perot cavity},''
  \href{http://dx.doi.org/10.1140/epjc/s10052-015-3869-8}{{\em Eur. Phys. J. C}
  {\bfseries 76} no.~1, (2016) 24},
  \href{http://arxiv.org/abs/1510.08052}{{\ttfamily arXiv:1510.08052
  [physics.optics]}}.

\bibitem{SAPPHIRES:2021vkz}
{\bfseries SAPPHIRES} Collaboration, K.~Homma {\em et~al.}, ``{Search for
  sub-eV axion-like resonance states via stimulated quasi-parallel laser
  collisions with the parameterization including fully asymmetric collisional
  geometry},'' \href{http://dx.doi.org/10.1007/JHEP12(2021)108}{{\em JHEP}
  {\bfseries 12} (2021) 108}, \href{http://arxiv.org/abs/2105.01224}{{\ttfamily
  arXiv:2105.01224 [hep-ex]}}.

\bibitem{Ortiz:2020tgs}
M.~D. Ortiz {\em et~al.}, ``{Design of the ALPS II optical system},''
  \href{http://dx.doi.org/10.1016/j.dark.2022.100968}{{\em Phys. Dark Univ.}
  {\bfseries 35} (2022) 100968},
  \href{http://arxiv.org/abs/2009.14294}{{\ttfamily arXiv:2009.14294
  [physics.optics]}}.

\bibitem{Lucente:2020whw}
G.~Lucente, P.~Carenza, T.~Fischer, M.~Giannotti, and A.~Mirizzi, ``{Heavy
  axion-like particles and core-collapse supernovae: constraints and impact on
  the explosion mechanism},''
  \href{http://dx.doi.org/10.1088/1475-7516/2020/12/008}{{\em JCAP} {\bfseries
  12} (2020) 008}, \href{http://arxiv.org/abs/2008.04918}{{\ttfamily
  arXiv:2008.04918 [hep-ph]}}.

\bibitem{Caputo:2022mah}
A.~Caputo, H.-T. Janka, G.~Raffelt, and E.~Vitagliano, ``{Low-Energy Supernovae
  Severely Constrain Radiative Particle Decays},''
  \href{http://arxiv.org/abs/2201.09890}{{\ttfamily arXiv:2201.09890
  [astro-ph.HE]}}.

\bibitem{Xiao:2020pra}
M.~Xiao, K.~M. Perez, M.~Giannotti, O.~Straniero, A.~Mirizzi, B.~W.
  Grefenstette, B.~M. Roach, and M.~Nynka, ``{Constraints on Axionlike
  Particles from a Hard X-Ray Observation of Betelgeuse},''
  \href{http://dx.doi.org/10.1103/PhysRevLett.126.031101}{{\em Phys. Rev.
  Lett.} {\bfseries 126} no.~3, (2021) 031101},
  \href{http://arxiv.org/abs/2009.09059}{{\ttfamily arXiv:2009.09059
  [astro-ph.HE]}}.

\bibitem{Keller:2021zbl}
A.~Keller, S.~O'Brien, A.~Kamdar, N.~M. Rapidis, A.~F. Leder, and K.~van
  Bibber, ``{A Model-independent Radio Telescope Dark Matter Search},''
  \href{http://dx.doi.org/10.3847/1538-4357/ac4d93}{{\em Astrophys. J.}
  {\bfseries 927} no.~1, (2022) 71},
  \href{http://arxiv.org/abs/2112.03439}{{\ttfamily arXiv:2112.03439
  [astro-ph.CO]}}.

\bibitem{Foster:2022fxn}
J.~W. Foster, S.~J. Witte, M.~Lawson, T.~Linden, V.~Gajjar, C.~Weniger, and
  B.~R. Safdi, ``{Extraterrestrial Axion Search with the Breakthrough Listen
  Galactic Center Survey},'' \href{http://arxiv.org/abs/2202.08274}{{\ttfamily
  arXiv:2202.08274 [astro-ph.CO]}}.

\bibitem{Chan:2021gjl}
M.~H. Chan, ``{Constraining the axion\textendash{}photon coupling using radio
  data of the Bullet cluster},''
  \href{http://dx.doi.org/10.1038/s41598-021-99495-3}{{\em Sci. Rep.}
  {\bfseries 11} no.~1, (2021) 20087},
  \href{http://arxiv.org/abs/2109.11734}{{\ttfamily arXiv:2109.11734
  [astro-ph.CO]}}.

\bibitem{Wouters:2013hua}
D.~Wouters and P.~Brun, ``{Constraints on Axion-like Particles from X-Ray
  Observations of the Hydra Galaxy Cluster},''
  \href{http://dx.doi.org/10.1088/0004-637X/772/1/44}{{\em Astrophys. J.}
  {\bfseries 772} (2013) 44}, \href{http://arxiv.org/abs/1304.0989}{{\ttfamily
  arXiv:1304.0989 [astro-ph.HE]}}.

\bibitem{Marsh:2017yvc}
M.~C.~D. Marsh, H.~R. Russell, A.~C. Fabian, B.~P. McNamara, P.~Nulsen, and
  C.~S. Reynolds, ``{A New Bound on Axion-Like Particles},''
  \href{http://dx.doi.org/10.1088/1475-7516/2017/12/036}{{\em JCAP} {\bfseries
  12} (2017) 036}, \href{http://arxiv.org/abs/1703.07354}{{\ttfamily
  arXiv:1703.07354 [hep-ph]}}.

\bibitem{Reynolds:2019uqt}
C.~S. Reynolds, M.~C.~D. Marsh, H.~R. Russell, A.~C. Fabian, R.~Smith,
  F.~Tombesi, and S.~Veilleux, ``{Astrophysical limits on very light axion-like
  particles from Chandra grating spectroscopy of NGC 1275},''
  \href{http://arxiv.org/abs/1907.05475}{{\ttfamily arXiv:1907.05475
  [hep-ph]}}.

\bibitem{Reynes:2021bpe}
J.~S. Reyn\'es, J.~H. Matthews, C.~S. Reynolds, H.~R. Russell, R.~N. Smith, and
  M.~C.~D. Marsh, ``{New constraints on light axion-like particles using
  Chandra transmission grating spectroscopy of the powerful cluster-hosted
  quasar H1821+643},'' \href{http://dx.doi.org/10.1093/mnras/stab3464}{{\em
  Mon. Not. Roy. Astron. Soc.} {\bfseries 510} no.~1, (2021) 1264--1277},
  \href{http://arxiv.org/abs/2109.03261}{{\ttfamily arXiv:2109.03261
  [astro-ph.HE]}}.

\bibitem{Bolliet:2020ofj}
B.~Bolliet, J.~Chluba, and R.~Battye, ``{Spectral distortion constraints on
  photon injection from low-mass decaying particles},''
  \href{http://dx.doi.org/10.1093/mnras/stab1997}{{\em Mon. Not. Roy. Astron.
  Soc.} {\bfseries 507} no.~3, (2021) 3148--3178},
  \href{http://arxiv.org/abs/2012.07292}{{\ttfamily arXiv:2012.07292
  [astro-ph.CO]}}.

\bibitem{Calore:2021hhn}
F.~Calore, P.~Carenza, C.~Eckner, T.~Fischer, M.~Giannotti, J.~Jaeckel,
  K.~Kotake, T.~Kuroda, A.~Mirizzi, and F.~Sivo, ``{3D template-based Fermi-LAT
  constraints on the diffuse supernova axion-like particle background},''
  \href{http://dx.doi.org/10.1103/PhysRevD.105.063028}{{\em Phys. Rev. D}
  {\bfseries 105} no.~6, (2022) 063028},
  \href{http://arxiv.org/abs/2110.03679}{{\ttfamily arXiv:2110.03679
  [astro-ph.HE]}}.

\bibitem{Calore:2020tjw}
F.~Calore, P.~Carenza, M.~Giannotti, J.~Jaeckel, and A.~Mirizzi, ``{Bounds on
  axionlike particles from the diffuse supernova flux},''
  \href{http://dx.doi.org/10.1103/PhysRevD.102.123005}{{\em Phys. Rev. D}
  {\bfseries 102} no.~12, (2020) 123005},
  \href{http://arxiv.org/abs/2008.11741}{{\ttfamily arXiv:2008.11741
  [hep-ph]}}.

\bibitem{Buen-Abad:2020zbd}
M.~A. Buen-Abad, J.~Fan, and C.~Sun, ``{Constraints on axions from cosmic
  distance measurements},''
  \href{http://dx.doi.org/10.1007/JHEP02(2022)103}{{\em JHEP} {\bfseries 02}
  (2022) 103}, \href{http://arxiv.org/abs/2011.05993}{{\ttfamily
  arXiv:2011.05993 [hep-ph]}}.

\bibitem{Fermi-LAT:2016nkz}
{\bfseries Fermi-LAT} Collaboration, M.~Ajello {\em et~al.}, ``{Search for
  Spectral Irregularities due to Photon\textendash{}Axionlike-Particle
  Oscillations with the Fermi Large Area Telescope},''
  \href{http://dx.doi.org/10.1103/PhysRevLett.116.161101}{{\em Phys. Rev.
  Lett.} {\bfseries 116} no.~16, (2016) 161101},
  \href{http://arxiv.org/abs/1603.06978}{{\ttfamily arXiv:1603.06978
  [astro-ph.HE]}}.

\bibitem{Meyer:2020vzy}
M.~Meyer and T.~Petrushevska, ``{Search for Axionlike-Particle-Induced Prompt
  $\gamma$-Ray Emission from Extragalactic Core-Collapse Supernovae with the
  $Fermi$ Large Area Telescope},''
  \href{http://dx.doi.org/10.1103/PhysRevLett.124.231101}{{\em Phys. Rev.
  Lett.} {\bfseries 124} no.~23, (2020) 231101},
  \href{http://arxiv.org/abs/2006.06722}{{\ttfamily arXiv:2006.06722
  [astro-ph.HE]}}. [Erratum: Phys.Rev.Lett. 125, 119901 (2020)].

\bibitem{Jacobsen:2022swa}
S.~Jacobsen, T.~Linden, and K.~Freese, ``{Constraining Axion-Like Particles
  with HAWC Observations of TeV Blazars},''
  \href{http://arxiv.org/abs/2203.04332}{{\ttfamily arXiv:2203.04332
  [hep-ph]}}.

\bibitem{HESS:2013udx}
{\bfseries H.E.S.S.} Collaboration, A.~Abramowski {\em et~al.}, ``{Constraints
  on axionlike particles with H.E.S.S. from the irregularity of the PKS
  2155-304 energy spectrum},''
  \href{http://dx.doi.org/10.1103/PhysRevD.88.102003}{{\em Phys. Rev. D}
  {\bfseries 88} no.~10, (2013) 102003},
  \href{http://arxiv.org/abs/1311.3148}{{\ttfamily arXiv:1311.3148
  [astro-ph.HE]}}.

\bibitem{Ayala:2014pea}
A.~Ayala, I.~Dom\'\i{}nguez, M.~Giannotti, A.~Mirizzi, and O.~Straniero,
  ``{Revisiting the bound on axion-photon coupling from Globular Clusters},''
  \href{http://dx.doi.org/10.1103/PhysRevLett.113.191302}{{\em Phys. Rev.
  Lett.} {\bfseries 113} no.~19, (2014) 191302},
  \href{http://arxiv.org/abs/1406.6053}{{\ttfamily arXiv:1406.6053
  [astro-ph.SR]}}.

\bibitem{Wadekar:2021qae}
D.~Wadekar and Z.~Wang, ``{Strong constraints on decay and annihilation of dark
  matter from heating of gas-rich dwarf galaxies},''
  \href{http://arxiv.org/abs/2111.08025}{{\ttfamily arXiv:2111.08025
  [hep-ph]}}.

\bibitem{Dessert:2022yqq}
C.~Dessert, D.~Dunsky, and B.~R. Safdi, ``{Upper limit on the axion-photon
  coupling from magnetic white dwarf polarization},''
  \href{http://arxiv.org/abs/2203.04319}{{\ttfamily arXiv:2203.04319
  [hep-ph]}}.

\bibitem{Li:2020pcn}
H.-J. Li, J.-G. Guo, X.-J. Bi, S.-J. Lin, and P.-F. Yin, ``{Limits on
  axion-like particles from Mrk 421 with 4.5-year period observations by
  ARGO-YBJ and Fermi-LAT},''
  \href{http://dx.doi.org/10.1103/PhysRevD.103.083003}{{\em Phys. Rev. D}
  {\bfseries 103} no.~8, (2021) 083003},
  \href{http://arxiv.org/abs/2008.09464}{{\ttfamily arXiv:2008.09464
  [astro-ph.HE]}}.

\bibitem{Foster:2020pgt}
J.~W. Foster, Y.~Kahn, O.~Macias, Z.~Sun, R.~P. Eatough, V.~I. Kondratiev,
  W.~M. Peters, C.~Weniger, and B.~R. Safdi, ``{Green Bank and Effelsberg Radio
  Telescope Searches for Axion Dark Matter Conversion in Neutron Star
  Magnetospheres},''
  \href{http://dx.doi.org/10.1103/PhysRevLett.125.171301}{{\em Phys. Rev.
  Lett.} {\bfseries 125} no.~17, (2020) 171301},
  \href{http://arxiv.org/abs/2004.00011}{{\ttfamily arXiv:2004.00011
  [astro-ph.CO]}}.

\bibitem{Darling:2020uyo}
J.~Darling, ``{New Limits on Axionic Dark Matter from the Magnetar PSR
  J1745-2900},'' \href{http://dx.doi.org/10.3847/2041-8213/abb23f}{{\em
  Astrophys. J. Lett.} {\bfseries 900} no.~2, (2020) L28},
  \href{http://arxiv.org/abs/2008.11188}{{\ttfamily arXiv:2008.11188
  [astro-ph.CO]}}.

\bibitem{Battye:2021yue}
R.~A. Battye, J.~Darling, J.~I. McDonald, and S.~Srinivasan, ``{Towards robust
  constraints on axion dark matter using PSR J1745-2900},''
  \href{http://dx.doi.org/10.1103/PhysRevD.105.L021305}{{\em Phys. Rev. D}
  {\bfseries 105} no.~2, (2022) L021305},
  \href{http://arxiv.org/abs/2107.01225}{{\ttfamily arXiv:2107.01225
  [astro-ph.CO]}}.

\bibitem{Vinyoles2015}
N.~{Vinyoles}, A.~{Serenelli}, F.~L. {Villante}, S.~{Basu}, J.~{Redondo}, and
  J.~{Isern}, ``{New axion and hidden photon constraints from a solar data
  global fit},'' \href{http://dx.doi.org/10.1088/1475-7516/2015/10/015}{{\em
  JCAP} {\bfseries 2015} no.~10, (Oct., 2015) 015},
  \href{http://arxiv.org/abs/1501.01639}{{\ttfamily arXiv:1501.01639
  [astro-ph.SR]}}.

\bibitem{Jaeckel:2017tud}
J.~Jaeckel, P.~C. Malta, and J.~Redondo, ``{Decay photons from the axionlike
  particles burst of type II supernovae},''
  \href{http://dx.doi.org/10.1103/PhysRevD.98.055032}{{\em Phys. Rev. D}
  {\bfseries 98} no.~5, (2018) 055032},
  \href{http://arxiv.org/abs/1702.02964}{{\ttfamily arXiv:1702.02964
  [hep-ph]}}.

\bibitem{Payez:2014xsa}
A.~Payez, C.~Evoli, T.~Fischer, M.~Giannotti, A.~Mirizzi, and A.~Ringwald,
  ``{Revisiting the SN1987A gamma-ray limit on ultralight axion-like
  particles},'' \href{http://dx.doi.org/10.1088/1475-7516/2015/02/006}{{\em
  JCAP} {\bfseries 02} (2015) 006},
  \href{http://arxiv.org/abs/1410.3747}{{\ttfamily arXiv:1410.3747
  [astro-ph.HE]}}.

\bibitem{Dessert:2020lil}
C.~Dessert, J.~W. Foster, and B.~R. Safdi, ``{X-ray Searches for Axions from
  Super Star Clusters},''
  \href{http://dx.doi.org/10.1103/PhysRevLett.125.261102}{{\em Phys. Rev.
  Lett.} {\bfseries 125} no.~26, (2020) 261102},
  \href{http://arxiv.org/abs/2008.03305}{{\ttfamily arXiv:2008.03305
  [hep-ph]}}.

\bibitem{Blout:2000uc}
B.~D. Blout, E.~J. Daw, M.~P. Decowski, P.~T.~P. Ho, L.~J. Rosenberg, and D.~B.
  Yu, ``{A Radio telescope search for axions},''
  \href{http://dx.doi.org/10.1086/318310}{{\em Astrophys. J.} {\bfseries 546}
  (2001) 825--828}, \href{http://arxiv.org/abs/astro-ph/0006310}{{\ttfamily
  arXiv:astro-ph/0006310}}.

\bibitem{Regis:2020fhw}
M.~Regis, M.~Taoso, D.~Vaz, J.~Brinchmann, S.~L. Zoutendijk, N.~F. Bouch\'e,
  and M.~Steinmetz, ``{Searching for light in the darkness: Bounds on ALP dark
  matter with the optical MUSE-faint survey},''
  \href{http://dx.doi.org/10.1016/j.physletb.2021.136075}{{\em Phys. Lett. B}
  {\bfseries 814} (2021) 136075},
  \href{http://arxiv.org/abs/2009.01310}{{\ttfamily arXiv:2009.01310
  [astro-ph.CO]}}.

\bibitem{Grin:2006aw}
D.~Grin, G.~Covone, J.-P. Kneib, M.~Kamionkowski, A.~Blain, and E.~Jullo, ``{A
  Telescope Search for Decaying Relic Axions},''
  \href{http://dx.doi.org/10.1103/PhysRevD.75.105018}{{\em Phys. Rev. D}
  {\bfseries 75} (2007) 105018},
  \href{http://arxiv.org/abs/astro-ph/0611502}{{\ttfamily
  arXiv:astro-ph/0611502}}.

\bibitem{Meyer:2016wrm}
M.~Meyer, M.~Giannotti, A.~Mirizzi, J.~Conrad, and M.~A. S\'anchez-Conde,
  ``{Fermi Large Area Telescope as a Galactic Supernovae Axionscope},''
  \href{http://dx.doi.org/10.1103/PhysRevLett.118.011103}{{\em Phys. Rev.
  Lett.} {\bfseries 118} no.~1, (2017) 011103},
  \href{http://arxiv.org/abs/1609.02350}{{\ttfamily arXiv:1609.02350
  [astro-ph.HE]}}.

\bibitem{Thorpe-Morgan:2020rwc}
C.~Thorpe-Morgan, D.~Malyshev, A.~Santangelo, J.~Jochum, B.~J\"ager, M.~Sasaki,
  and S.~Saeedi, ``{THESEUS insights into axionlike particles, dark photon, and
  sterile neutrino dark matter},''
  \href{http://dx.doi.org/10.1103/PhysRevD.102.123003}{{\em Phys. Rev. D}
  {\bfseries 102} no.~12, (2020) 123003},
  \href{http://arxiv.org/abs/2008.08306}{{\ttfamily arXiv:2008.08306
  [astro-ph.HE]}}.

\bibitem{Dekker:2021bos}
A.~Dekker, E.~Peerbooms, F.~Zimmer, K.~C.~Y. Ng, and S.~Ando, ``{Searches for
  sterile neutrinos and axionlike particles from the Galactic halo with
  eROSITA},'' \href{http://dx.doi.org/10.1103/PhysRevD.104.023021}{{\em Phys.
  Rev. D} {\bfseries 104} no.~2, (2021) 023021},
  \href{http://arxiv.org/abs/2103.13241}{{\ttfamily arXiv:2103.13241
  [astro-ph.HE]}}.

\bibitem{Dolan:2021rya}
M.~J. Dolan, F.~J. Hiskens, and R.~R. Volkas, ``{Constraining axion-like
  particles using the white dwarf initial-final mass relation},''
  \href{http://dx.doi.org/10.1088/1475-7516/2021/09/010}{{\em JCAP} {\bfseries
  09} (2021) 010}, \href{http://arxiv.org/abs/2102.00379}{{\ttfamily
  arXiv:2102.00379 [hep-ph]}}.

\bibitem{Foster:2021ngm}
J.~W. Foster, M.~Kongsore, C.~Dessert, Y.~Park, N.~L. Rodd, K.~Cranmer, and
  B.~R. Safdi, ``{Deep Search for Decaying Dark Matter with XMM-Newton
  Blank-Sky Observations},''
  \href{http://dx.doi.org/10.1103/PhysRevLett.127.051101}{{\em Phys. Rev.
  Lett.} {\bfseries 127} no.~5, (2021) 051101},
  \href{http://arxiv.org/abs/2102.02207}{{\ttfamily arXiv:2102.02207
  [astro-ph.CO]}}.

\bibitem{Cadamuro:2011fd}
D.~Cadamuro and J.~Redondo, ``{Cosmological bounds on pseudo Nambu-Goldstone
  bosons},'' \href{http://dx.doi.org/10.1088/1475-7516/2012/02/032}{{\em JCAP}
  {\bfseries 02} (2012) 032}, \href{http://arxiv.org/abs/1110.2895}{{\ttfamily
  arXiv:1110.2895 [hep-ph]}}.

\bibitem{NOMAD:2000usb}
{\bfseries NOMAD} Collaboration, P.~Astier {\em et~al.}, ``{Search for eV
  (pseudo)scalar penetrating particles in the SPS neutrino beam},''
  \href{http://dx.doi.org/10.1016/S0370-2693(00)00375-0}{{\em Phys. Lett. B}
  {\bfseries 479} (2000) 371--380}.

\bibitem{Jaeckel:2015jla}
J.~Jaeckel and M.~Spannowsky, ``{Probing MeV to 90 GeV axion-like particles
  with LEP and LHC},''
  \href{http://dx.doi.org/10.1016/j.physletb.2015.12.037}{{\em Phys. Lett.}
  {\bfseries B753} (2016) 482--487},
\href{http://arxiv.org/abs/1509.00476}{{\ttfamily arXiv:1509.00476 [hep-ph]}}.

\bibitem{DELPHI:1991emv}
{\bfseries DELPHI} Collaboration, P.~Abreu {\em et~al.}, ``{The reaction e+ e-
  ---\ensuremath{>} gamma gamma (gamma) at Z0 energies},''
  \href{http://dx.doi.org/10.1016/0370-2693(91)90819-C}{{\em Phys. Lett. B}
  {\bfseries 268} (1991) 296--304}.

\bibitem{DELPHI:1994mra}
{\bfseries DELPHI} Collaboration, P.~Abreu {\em et~al.}, ``{Measurement of the
  e+ e- ---\ensuremath{>} gamma gamma (gamma) cross-section at LEP energies},''
  \href{http://dx.doi.org/10.1016/0370-2693(94)90745-5}{{\em Phys. Lett. B}
  {\bfseries 327} (1994) 386--396}.

\bibitem{L3:1994shn}
{\bfseries L3} Collaboration, M.~Acciarri {\em et~al.}, ``{Search for anomalous
  Z --\ensuremath{>} gamma gamma gamma events at LEP},''
  \href{http://dx.doi.org/10.1016/0370-2693(95)01612-T}{{\em Phys. Lett. B}
  {\bfseries 345} (1995) 609--616}.

\bibitem{L3:1995nbq}
{\bfseries L3} Collaboration, M.~Acciarri {\em et~al.}, ``{Tests of QED at LEP
  energies using e+ e- --\ensuremath{>} gamma gamma (gamma) and e+ e-
  --\ensuremath{>} lepton+ lepton- gamma gamma},''
  \href{http://dx.doi.org/10.1016/0370-2693(95)00527-R}{{\em Phys. Lett. B}
  {\bfseries 353} (1995) 136--144}.

\bibitem{CHARM:1985anb}
{\bfseries CHARM} Collaboration, F.~Bergsma {\em et~al.}, ``{Search for Axion
  Like Particle Production in 400-{GeV} Proton - Copper Interactions},''
  \href{http://dx.doi.org/10.1016/0370-2693(85)90400-9}{{\em Phys. Lett. B}
  {\bfseries 157} (1985) 458--462}.

\bibitem{Riordan:1987aw}
E.~M. Riordan {\em et~al.}, ``{A Search for Short Lived Axions in an Electron
  Beam Dump Experiment},''
\href{http://dx.doi.org/10.1103/PhysRevLett.59.755}{{\em Phys. Rev. Lett.}
  {\bfseries 59} (1987) 755}.

\bibitem{Dobrich:2017gcm}
B.~D\"obrich, ``{Axion-like Particles from Primakov production in
  beam-dumps},''
  \href{http://dx.doi.org/10.23727/CERN-Proceedings-2018-001.253}{{\em CERN
  Proc.} {\bfseries 1} (2018) 253},
  \href{http://arxiv.org/abs/1708.05776}{{\ttfamily arXiv:1708.05776
  [hep-ph]}}.

\bibitem{Bjorken:1988as}
J.~D. Bjorken, S.~Ecklund, W.~R. Nelson, A.~Abashian, C.~Church, B.~Lu, L.~W.
  Mo, T.~A. Nunamaker, and P.~Rassmann, ``{Search for Neutral Metastable
  Penetrating Particles Produced in the SLAC Beam Dump},''
\href{http://dx.doi.org/10.1103/PhysRevD.38.3375}{{\em Phys. Rev.} {\bfseries
  D38} (1988) 3375}.

\bibitem{Dolan:2017osp}
M.~J. Dolan, T.~Ferber, C.~Hearty, F.~Kahlhoefer, and K.~Schmidt-Hoberg,
  ``{Revised constraints and Belle II sensitivity for visible and invisible
  axion-like particles},''
  \href{http://dx.doi.org/10.1007/JHEP12(2017)094}{{\em JHEP} {\bfseries 12}
  (2017) 094}, \href{http://arxiv.org/abs/1709.00009}{{\ttfamily
  arXiv:1709.00009 [hep-ph]}}. [Erratum: JHEP 03, 190 (2021)].

\bibitem{Blumlein:1990ay}
J.~Blumlein {\em et~al.}, ``{Limits on neutral light scalar and pseudoscalar
  particles in a proton beam dump experiment},''
  \href{http://dx.doi.org/10.1007/BF01548556}{{\em Z. Phys. C} {\bfseries 51}
  (1991) 341--350}.

\bibitem{NA64:2020qwq}
{\bfseries NA64} Collaboration, D.~Banerjee {\em et~al.}, ``{Search for
  Axionlike and Scalar Particles with the NA64 Experiment},''
  \href{http://dx.doi.org/10.1103/PhysRevLett.125.081801}{{\em Phys. Rev.
  Lett.} {\bfseries 125} no.~8, (2020) 081801},
  \href{http://arxiv.org/abs/2005.02710}{{\ttfamily arXiv:2005.02710
  [hep-ex]}}.

\bibitem{Belle-II:2020jti}
{\bfseries Belle-II} Collaboration, F.~Abudin\'en {\em et~al.}, ``{Search for
  Axion-Like Particles produced in $e^+e^-$ collisions at Belle II},''
  \href{http://dx.doi.org/10.1103/PhysRevLett.125.161806}{{\em Phys. Rev.
  Lett.} {\bfseries 125} no.~16, (2020) 161806},
  \href{http://arxiv.org/abs/2007.13071}{{\ttfamily arXiv:2007.13071
  [hep-ex]}}.

\bibitem{Aloni:2019ruo}
D.~Aloni, C.~Fanelli, Y.~Soreq, and M.~Williams, ``{Photoproduction of
  Axionlike Particles},''
  \href{http://dx.doi.org/10.1103/PhysRevLett.123.071801}{{\em Phys. Rev.
  Lett.} {\bfseries 123} no.~7, (2019) 071801},
  \href{http://arxiv.org/abs/1903.03586}{{\ttfamily arXiv:1903.03586
  [hep-ph]}}.

\bibitem{Blumlein:1991xh}
J.~Blumlein {\em et~al.}, ``{Limits on the mass of light (pseudo)scalar
  particles from Bethe-Heitler e+ e- and mu+ mu- pair production in a proton -
  iron beam dump experiment},''
  \href{http://dx.doi.org/10.1142/S0217751X9200171X}{{\em Int. J. Mod. Phys. A}
  {\bfseries 7} (1992) 3835--3850}.

\bibitem{Fukuda:2022not}
H.~Fukuda, H.~Otono, and S.~Shirai, ``{QCD Axion Search with ILC Beam
  Facility},'' \href{http://arxiv.org/abs/2203.06137}{{\ttfamily
  arXiv:2203.06137 [hep-ph]}}.

\bibitem{Behnke:2013xla}
T.~Behnke, J.~E. Brau, B.~Foster, J.~Fuster, M.~Harrison, J.~M. Paterson,
  M.~Peskin, M.~Stanitzki, N.~Walker, and H.~Yamamoto, ``{The International
  Linear Collider Technical Design Report - Volume 1: Executive Summary},''
  \href{http://arxiv.org/abs/1306.6327}{{\ttfamily arXiv:1306.6327
  [physics.acc-ph]}}.

\bibitem{Bonivento:2019sri}
W.~M. Bonivento, D.~Kim, and K.~Sinha, ``{PASSAT: Particle Accelerator
  helioScopes for Slim Axion-like-particle deTection},''
  \href{http://dx.doi.org/10.1140/epjc/s10052-020-7719-y}{{\em Eur. Phys. J. C}
  {\bfseries 80} no.~2, (2020) 164},
  \href{http://arxiv.org/abs/1909.03071}{{\ttfamily arXiv:1909.03071
  [hep-ph]}}.

\bibitem{Dev:2021ofc}
P.~S.~B. Dev, D.~Kim, K.~Sinha, and Y.~Zhang, ``{PASSAT at future neutrino
  experiments: Hybrid beam-dump-helioscope facilities to probe light axionlike
  particles},'' \href{http://dx.doi.org/10.1103/PhysRevD.104.035037}{{\em Phys.
  Rev. D} {\bfseries 104} no.~3, (2021) 035037},
  \href{http://arxiv.org/abs/2101.08781}{{\ttfamily arXiv:2101.08781
  [hep-ph]}}.

\bibitem{Dent:2019ueq}
J.~B. Dent, B.~Dutta, D.~Kim, S.~Liao, R.~Mahapatra, K.~Sinha, and A.~Thompson,
  ``{New Directions for Axion Searches via Scattering at Reactor Neutrino
  Experiments},'' \href{http://dx.doi.org/10.1103/PhysRevLett.124.211804}{{\em
  Phys. Rev. Lett.} {\bfseries 124} no.~21, (2020) 211804},
  \href{http://arxiv.org/abs/1912.05733}{{\ttfamily arXiv:1912.05733
  [hep-ph]}}.

\bibitem{Beacham:2019nyx}
J.~Beacham {\em et~al.}, ``{Physics Beyond Colliders at CERN: Beyond the
  Standard Model Working Group Report},''
  \href{http://dx.doi.org/10.1088/1361-6471/ab4cd2}{{\em J. Phys. G} {\bfseries
  47} no.~1, (2020) 010501}, \href{http://arxiv.org/abs/1901.09966}{{\ttfamily
  arXiv:1901.09966 [hep-ex]}}.

\bibitem{Georgi:1986df}
H.~Georgi, D.~B. Kaplan, and L.~Randall, ``{Manifesting the Invisible Axion at
  Low-energies},''
\href{http://dx.doi.org/10.1016/0370-2693(86)90688-X}{{\em Phys. Lett.}
  {\bfseries 169B} (1986) 73--78}.

\bibitem{DiLuzio:2016sbl}
L.~Di~Luzio, F.~Mescia, and E.~Nardi, ``{Redefining the Axion Window},''
  \href{http://dx.doi.org/10.1103/PhysRevLett.118.031801}{{\em Phys. Rev.
  Lett.} {\bfseries 118} no.~3, (2017) 031801},
\href{http://arxiv.org/abs/1610.07593}{{\ttfamily arXiv:1610.07593 [hep-ph]}}.

\bibitem{DiLuzio:2017pfr}
L.~Di~Luzio, F.~Mescia, and E.~Nardi, ``{Window for preferred axion models},''
  \href{http://dx.doi.org/10.1103/PhysRevD.96.075003}{{\em Phys. Rev.}
  {\bfseries D96} no.~7, (2017) 075003},
  \href{http://arxiv.org/abs/1705.05370}{{\ttfamily 1705.05370}}.

\bibitem{Farina:2016tgd}
M.~Farina, D.~Pappadopulo, F.~Rompineve, and A.~Tesi, ``{The photo-philic QCD
  axion},'' \href{http://dx.doi.org/10.1007/JHEP01(2017)095}{{\em JHEP}
  {\bfseries 01} (2017) 095},
\href{http://arxiv.org/abs/1611.09855}{{\ttfamily arXiv:1611.09855 [hep-ph]}}.

\bibitem{Darme:2020gyx}
L.~Darm\'e, L.~Di~Luzio, M.~Giannotti, and E.~Nardi, ``{Selective enhancement
  of the QCD axion couplings},''
  \href{http://arxiv.org/abs/2010.15846}{{\ttfamily arXiv:2010.15846
  [hep-ph]}}.

\bibitem{Agrawal:2021dbo}
P.~Agrawal {\em et~al.}, ``{Feebly-interacting particles: FIPs 2020 workshop
  report},'' \href{http://dx.doi.org/10.1140/epjc/s10052-021-09703-7}{{\em Eur.
  Phys. J. C} {\bfseries 81} no.~11, (2021) 1015},
  \href{http://arxiv.org/abs/2102.12143}{{\ttfamily arXiv:2102.12143
  [hep-ph]}}.

\bibitem{Hook:2018jle}
A.~Hook, ``{Solving the Hierarchy Problem Discretely},''
  \href{http://dx.doi.org/10.1103/PhysRevLett.120.261802}{{\em Phys. Rev.
  Lett.} {\bfseries 120} no.~26, (2018) 261802},
  \href{http://arxiv.org/abs/1802.10093}{{\ttfamily arXiv:1802.10093
  [hep-ph]}}.

\bibitem{DiLuzio:2021pxd}
L.~Di~Luzio, B.~Gavela, P.~Quilez, and A.~Ringwald, ``{An even lighter QCD
  axion},'' \href{http://dx.doi.org/10.1007/JHEP05(2021)184}{{\em JHEP}
  {\bfseries 05} (2021) 184}, \href{http://arxiv.org/abs/2102.00012}{{\ttfamily
  arXiv:2102.00012 [hep-ph]}}.

\bibitem{DiLuzio:2021gos}
L.~Di~Luzio, B.~Gavela, P.~Quilez, and A.~Ringwald, ``{Dark matter from an even
  lighter QCD axion: trapped misalignment},''
  \href{http://dx.doi.org/10.1088/1475-7516/2021/10/001}{{\em JCAP} {\bfseries
  10} (2021) 001}, \href{http://arxiv.org/abs/2102.01082}{{\ttfamily
  arXiv:2102.01082 [hep-ph]}}.

\bibitem{Dafni:2018tvj}
T.~Dafni, C.~A.~J. O'Hare, B.~Laki\'c, J.~Gal\'an, F.~J. Iguaz, I.~G.
  Irastorza, K.~Jakov\v{c}ic, G.~Luz\'on, J.~Redondo, and E.~Ruiz~Ch\'oliz,
  ``{Weighing the solar axion},''
  \href{http://dx.doi.org/10.1103/PhysRevD.99.035037}{{\em Phys. Rev. D}
  {\bfseries 99} no.~3, (2019) 035037},
  \href{http://arxiv.org/abs/1811.09290}{{\ttfamily arXiv:1811.09290
  [hep-ph]}}.

\bibitem{Jaeckel:2018mbn}
J.~Jaeckel and L.~J. Thormaehlen, ``{Distinguishing Axion Models with IAXO},''
  \href{http://dx.doi.org/10.1088/1475-7516/2019/03/039}{{\em JCAP} {\bfseries
  03} (2019) 039}, \href{http://arxiv.org/abs/1811.09278}{{\ttfamily
  arXiv:1811.09278 [hep-ph]}}.

\bibitem{Gori:2020xvq}
S.~Gori, G.~Perez, and K.~Tobioka, ``{KOTO vs. NA62 Dark Scalar Searches},''
  \href{http://dx.doi.org/10.1007/JHEP08(2020)110}{{\em JHEP} {\bfseries 08}
  (2020) 110}, \href{http://arxiv.org/abs/2005.05170}{{\ttfamily
  arXiv:2005.05170 [hep-ph]}}.

\bibitem{Kling:2020mch}
F.~Kling and S.~Trojanowski, ``{Looking forward to test the KOTO anomaly with
  FASER},'' \href{http://dx.doi.org/10.1103/PhysRevD.102.015032}{{\em Phys.
  Rev. D} {\bfseries 102} no.~1, (2020) 015032},
  \href{http://arxiv.org/abs/2006.10630}{{\ttfamily arXiv:2006.10630
  [hep-ph]}}.

\bibitem{Arias:2010bh}
P.~Arias, J.~Jaeckel, J.~Redondo, and A.~Ringwald, ``{Optimizing
  Light-Shining-through-a-Wall Experiments for Axion and other WISP
  Searches},'' \href{http://dx.doi.org/10.1103/PhysRevD.82.115018}{{\em Phys.
  Rev. D} {\bfseries 82} (2010) 115018},
  \href{http://arxiv.org/abs/1009.4875}{{\ttfamily arXiv:1009.4875 [hep-ph]}}.

\bibitem{CAST:2013bqn}
{\bfseries CAST} Collaboration, M.~Arik {\em et~al.}, ``{Search for Solar
  Axions by the CERN Axion Solar Telescope with $^3$He Buffer Gas: Closing the
  Hot Dark Matter Gap},''
  \href{http://dx.doi.org/10.1103/PhysRevLett.112.091302}{{\em Phys. Rev.
  Lett.} {\bfseries 112} no.~9, (2014) 091302},
  \href{http://arxiv.org/abs/1307.1985}{{\ttfamily arXiv:1307.1985 [hep-ex]}}.

\bibitem{ALPS:2009des}
{\bfseries ALPS} Collaboration, K.~Ehret {\em et~al.}, ``{Resonant laser power
  build-up in ALPS: A 'Light-shining-through-walls' experiment},''
  \href{http://dx.doi.org/10.1016/j.nima.2009.10.102}{{\em Nucl. Instrum. Meth.
  A} {\bfseries 612} (2009) 83--96},
  \href{http://arxiv.org/abs/0905.4159}{{\ttfamily arXiv:0905.4159
  [physics.ins-det]}}.

\end{thebibliography}\endgroup
\end{document}